\documentclass{aa}
\usepackage{graphicx}
\usepackage{txfonts}
\usepackage{hyperref}
\usepackage[normalem]{ulem}
\hypersetup{colorlinks = true, citecolor = {blue}, urlcolor= {blue}}
\usepackage{longtable}
\usepackage{placeins}

\newcommand{\msun}{{\rm M}_{\odot}}
\newcommand{\lsun}{{\rm L}_{\odot}}

\def\gapprox{\;\rlap{\lower 3.0pt                       
        \hbox{$\sim$}}\raise 2.5pt\hbox{$>$}\;}
\def\lapprox{\;\rlap{\lower 3.1pt                       
        \hbox{$\sim$}}\raise 2.7pt\hbox{$<$}\;}

\usepackage{threeparttable}

\newcommand{\be}{ \begin{equation} }
\newcommand{\ee}{\end{equation}}

\newcommand{\ben}{\begin{enumerate}}
\newcommand{\een}{\end{enumerate}}
\renewcommand{\arraystretch}{1.2}

\renewenvironment{thebibliography}[1]{\begin{oldthebibliography}{#1}\setlength{\parskip}{0ex}\setlength{\itemsep}{0ex}}{\end{oldthebibliography}}

\usepackage{threeparttable}
\usepackage{diagbox}
\usepackage{siunitx}
\usepackage{ltablex}
\makeatletter
\renewcommand*\aa@pageof{, page \thepage{} of \pageref*{LastPage}}
\makeatother

\usepackage[dvipsnames]{xcolor}

\definecolor{darkgreen}{RGB}{31, 207, 31}

\usepackage{supertabular}

\usepackage{booktabs}

\newcommand{\rev}[1]{\textbf{\textcolor{red}{#1}}}

\usepackage{threeparttable}
\usepackage{multirow}
\usepackage{booktabs}
\newcommand{\mesa}{\textsc{mesa}}
\newcommand{\mesarsp}{\textsc{mesa-rsp}}
\defcitealias{kova24}{K24}
\defcitealias{marc15}{M15}

\begin{document}
    \title{Theoretical RR Lyrae Instability Strip from MESA-RSP Pulsation models and updated BaSTI Evolutionary Tracks}

    \author{M. Deka
          \inst{1},
          M. Marconi
          \inst{1},
          R. Molinaro
          \inst{1},
          A. Bhardwaj
          \inst{2},
          S. Cassisi
          \inst{3,4},
          E. Luongo
          \inst{1},
          G. De Somma
          \inst{1,5},
          I. Musella
          \inst{1}
          \and
          V. Ripepi
          \inst{1}
          }

   \institute{INAF-Osservatorio astronomico di Capodimonte, Via Moiariello 16, I-80131 Napoli, Italy\\
              \email{mami.deka@inaf.it}
              \and
              Inter-University Centre for Astronomy and Astrophysics (IUCAA), Post Bag 4, Ganeshkhind, Pune 411 007, India.
              \and
              INAF – Osservatorio Astronomico di Abruzzo, via M. Maggini snc, 64100 Teramo, Italy.
              \and
              INFN – Sezione di Pisa, Largo Pontecorvo 3, 56127 Pisa, Italy.
              \and
              Istituto Nazionale di Fisica Nucleare (INFN)—Sez. di Napoli, Compl. Univ. di Monte S. Angelo, Edificio G, Via Cinthia, I-80126, Napoli, Italy.
              }

   \date{}
	
\newcommand{\mar}[1]{\textcolor{blue}{#1}} 
\newcommand{\mami}[1]{\textcolor{magenta}{#1}}

\abstract
    {RR Lyraes (RRLs) are important tracers of old stellar populations and crucial distance indicators. To complement large variability surveys such as \textit{Gaia} and Rubin-LSST, the topology of the RRL instability strip (IS) and their period-luminosity-metallicity (PLZ) and period-Wesenheit-metallicity (PWZ) relations need to be updated using state-of-the-art stellar evolution and pulsation models.}
    {We employ the updated set of BaSTI horizontal-branch (HB) evolutionary models, with finer sampling in effective temperature and metallicity, and compute the corresponding pulsation models using the \mesarsp\ module building on the approach of Marconi et al. (2015). We aim to quantify how the predicted IS boundaries and PLZ/PWZ relations depend on the adopted convection prescription and mixing-length parameter.}
    {We compute a grid of linear non-adiabatic radial pulsation models with \mesarsp, covering $-2.2 \leq [\rm{Fe/H}] \leq +0.06$~dex. Seven convection configurations are explored including the recently empirically calibrated prescription for RRLs by Kovács et al. (2024). Analytical IS boundaries are derived for RRab and RRc pulsators and transformed into several photometric systems (Gaia, OGLE, 2MASS, and LSST) and the Kiel diagram.}
    {The updated evolutionary models do not substantially change the pulsation predictions: RRab IS edges agree well with previous results, while the RRc red edge is slightly redder. The width of the IS decreases systematically with increasing $\alpha_{\rm MLT}$, and RRc is relatively more sensitive to the convection prescription than RRab. Testing both linear and quadratic edge fits, RRc IS favours a quadratic luminosity term in most cases, indicating intrinsic curvature of the first-overtone IS. The observed sample lies well within our predicted IS edges when compared in the $(J-K{\rm s})_0$ versus $M_{K_{\rm s}}$ color–magnitude diagram. The PLZ and PWZ relations are also in good agreement with previous empirical and theoretical calibrations across the NIR bands; however, the convection prescription giving the closest match is wavelength dependent. The theoretical \textit{Gaia} PWZ relation shows a modest offset from the empirical Milky Way field relation, which diminishes significantly when compared at matching cluster metallicities.}
    {The predicted IS boundaries and PLZ/PWZ relations exhibit a coupled dependence on pulsation mode, metallicity, and convective treatment, with no single prescription performing uniformly best across all observational planes. The theoretical relations presented here are directly applicable to large RRL datasets from ongoing and forthcoming surveys, including \textit{Gaia} and Rubin-LSST.}

   \titlerunning{RR Lyrae Instability Strip edges}
   \authorrunning{M.\ Deka et al.}

   \maketitle
   
\section{Introduction}
\label{sec:intro}
RR Lyrae stars (RRLs) are low-mass Population~II pulsating variables, with masses of $0.5$--$0.8\,\msun$ and periods of $\sim0.2$--$1.0$~days \citep{smit95,cate09,marc15,bhar20,clem23}, occupying the classical instability strip (IS) at the intersection of the horizontal branch (HB) in the Hertzsprung-Russell diagram (HRD), \citep{cate04,cate09}. As tracers of ancient ($\gtrsim10$ Gyr) populations \citep{baad44,dora25}, RRLs are powerful probes of the assembly history of the Galactic halo, thick disk, bulge, and dwarf galaxies \citep[e.g.,][]{helm08,luon24,medi25b}. They are classified by pulsation mode: fundamental-mode (RRab, $P\sim0.29$--$1.00$ d, high-amplitude asymmetric light curves) and first-overtone (RRc, $P\sim0.20$--$0.51$ d, lower-amplitude sinusoidal curves), with double-mode RRd pulsators oscillating in both modes \citep{clem23}.

RRLs have long been recognized as fundamental distance indicators, owing to the well-established correlation between their absolute visual magnitude and iron abundance ([Fe/H]) \citep[e.g.,][]{cate04,marc15}. Since the pioneering work of \citet{long86}, their Period-Luminosity (PL) and Period-Wesenheit (PW) relations in the near-infrared (NIR), have been extensively investigated and continuously refined through both theoretical modeling and observational studies. These relations establish RRLs as robust tracers of distances within the Milky Way and nearby galaxies \citep{smit95, cate15, bhar23a,bhar24a,ngeo26}. The advent of large-scale photometric surveys such as the \textit{Gaia} mission \citep{gaia16, gaia23} and the  Rubin Observatory Legacy Survey of Space and Time \citep[LSST;][]{ivez19} and the forthcoming Nancy Grace Roman Space Telescope 
\citep[Roman;][]{akeson2019} has substantially increased and will further increase, the census of known RRLs while significantly improving the precision and homogeneity of their observed properties. Such high-quality datasets provide stringent empirical constraints on pulsation models and offer a valuable opportunity to refine theoretical predictions of IS boundaries and other model-dependent quantities.

A fundamental concept in the theory of pulsating stars is the IS, the region in the HRD, where stellar pulsations are excited by the $\kappa$ and $\gamma$-mechanisms operating in the Hydrogen and partial Helium ionization zones \citep{cox80,bono94}. The location and topology of the RRL IS depend sensitively on stellar parameters such as mass, luminosity, effective temperature, and chemical composition, as well as on the treatment of convection in pulsation models 
(\citet{cris04}, \citet[][hereafter \citetalias{marc15}]{marc15}). Precise theoretical predictions of the IS are essential for interpreting observed RRL populations and for deriving reliable PL and PW relations. Over the past decades, significant progress has been made in modeling RRL pulsations. Nonlinear convective pulsation models have provided detailed predictions of light curves, pulsation amplitudes, and IS boundaries \citep[e.g.,][]{bono94,koll00,koll02,szab04,marc15,marc22}. In particular, \citetalias{marc15}  presented a comprehensive theoretical framework based on pulsation models coupled with evolutionary constraints from Bag of Stellar Tracks and Isochrones (BaSTI) stellar models by \citet{Pietrinferni06}. Their work established widely used relations for the IS boundaries and provided a theoretical basis for interpreting RRL observations across a range of metallicities ($0.0001<Z<0.02$). Although these models have successfully reproduced many observed properties of RRLs, they were constructed using earlier generations of evolutionary calculations and a particular set of convection parameters. Recent advances in stellar evolution modeling \citep{Pietrinferni21} and alternative treatments of convection in \mesarsp\ provide the possibility to reassess and further improve these predictions.

Recent advances in stellar evolutionary calculations have led to updated predictions for HB stars \citep{hida18, Pietrinferni21}. In particular, the revised BaSTI database incorporates improved input physics, including updated solar metal mixtures, electron conduction opacities, and nuclear reaction rates, leading to revisions in the predicted properties of HB stars (for more detail see Sect.~\ref{sec:ev models}). While these updates are generally expected to preserve the overall agreement with previous calculations, they may introduce modest changes in the mass–luminosity distribution and, consequently, small but potentially relevant shifts in the predicted location of the RRL IS. Consequently, the theoretical relations should be reexamined in light of these updated evolutionary models.

Another important aspect that remains insufficiently explored is the sensitivity of the IS boundaries to the adopted convection parameters. Convection strongly influences the red edge of the IS, and uncertainties in convection modeling represent a major source of theoretical error in pulsation models \citep[e.g.,][]{stel82a,bono94,kova23,deka24,deka26}. 
For example \citet{cris04,cris04a}  showed that a higher value of the mixing length parameter is required to fit the position of the RRL IS red edge than for the blue edge and that adopting a fixed value for the whole IS might introduce errors. Moreover, the model fitting of multiband light curves \citep[see e.g.][]{mc05,mdi07} suggests that higher mixing length values are required to fit the pulsation amplitude and morphology of RRcs than RRabs. Recently, \citet{kova23,kova24} performed a detailed calibration of turbulent convection (TC) parameters for RRLs in the globular cluster M3, using both radial velocity (RV) and $V$-band light curves. However, this calibration was limited to narrow period ranges; $0.50-0.71$ days for RRab and $0.29-0.34$ days for RRc, so its applicability across the full RRL population has yet to be tested. While \citet{kova23, kova24} provided distinct sets of TC parameters for RRab and RRc stars (optimized separately for the Budapest-Florida and the Radial Stellar Pulsation (RSP) module of the Modules for Experiments in Stellar Astrophysics \citep[\mesa;][]{paxt10,paxt13,paxt15,paxt18,paxt19,jerm23} codes) and successfully reproduced the observed RV curves, notable discrepancies remain in the modeled light curves. Additional observational constraints have recently been provided by \citet{reye24}, who compiled a large catalog of RRLs in Galactic globular clusters using Gaia DR3 and derived empirical IS boundaries. When compared with theoretical predictions based on \mesa\ models, their analysis suggested that different helium abundances were required to reproduce the observed locations of RRab and RRc stars within the IS. However, since these theoretical boundaries were derived using a single set of TC parameters, the extent to which uncertainties in convection modeling may contribute to the observed discrepancies remains unclear \citep[see for example][]{somm22,deka24}. A systematic investigation of the effects of different mixing-length values, convection parameter sets, as well as chemical composition, is therefore essential to assess the robustness of theoretical IS predictions.
Furthermore, modern photometric surveys operate in specific photometric systems, such as those of \textit{Gaia}, LSST, Roman, and the James Webb Space Telescope (JWST). This requires theoretical predictions to be expressed directly in the corresponding observational planes, rather than solely in the HRD or traditional optical bands.

In this work, we present updated theoretical predictions for the RRL IS based on a new grid of pulsation models computed with the \mesarsp\ module and evolutionary constraints from updated BaSTI HB models. Our model grid spans a range of metallicities extending to slightly more metal-rich regimes than those considered by \citetalias{marc15}. We also explore two TC parameter sets in the \mesarsp. In addition, we explore the impact of different mixing-length values on the location of the IS boundaries and quantify the resulting shifts in effective temperature. Finally, we derive the predicted IS edges into the  Optical Gravitational Lensing Experiment (OGLE), \textit{Gaia}, Two Micron All Sky Survey (2MASS) and  LSST photometric systems, providing relations that can be directly compared with the observations.

This paper is organised as follows. Sect.~\ref{sec:ev models} 
describes the BaSTI evolutionary models. Sect.~\ref{sec:pulsation_model}--\ref{sec:pulsation_stability} 
present the \mesarsp\ pulsation models, the model grid, the 
convection parameter sets, and the model selection criteria. 
Sect.~\ref{sec:results} analyses the IS topology and its 
dependence on metallicity and convection prescription across 
multiple photometric systems, and presents comparisons with 
previous theoretical work and observational data. 
Section~\ref{sec:discussion} discusses the implications, caveats, 
and limitations of the present work. Sect.~\ref{sec:summary} 
summarises the main conclusions.

\section{Pulsation models}
\label{sec:data}
\subsection{Construction of the pulsation models grid based on BaSTI horizontal branch models}
\label{sec:ev models}
The pulsation properties of RRLs depend critically on their underlying evolutionary structure, particularly their mass–luminosity–metallicity and effective temperature distribution along the HB. We adopted evolutionary models from the updated BaSTI stellar evolution database \citep{hida18,Pietrinferni21}, which provides state-of-the-art calculations for low-mass stars across a wide range of chemical compositions. Compared to earlier versions, the updated BaSTI models incorporate revised input physics, including updated radiative opacities, nuclear reaction rates, and solar heavy-element abundances, improving predictions for the structure and evolution of HB stars.

We selected zero-age horizontal branch (ZAHB) models spanning a wide metallicity range, covering both metal-poor and metal-rich regimes relevant for RRL populations in the Galactic halo, thick disk, and bulge. The adopted metallicities correspond to iron abundances in the approximate range ($-2.2 \le \rm{[Fe/H]} \le +0.06$)~dex, assuming an $\alpha$-enhanced mixture of $[\alpha/\rm{Fe}] = +0.4$. The initial helium mass fraction $Y$ is calculated for a given $Z$ assuming primordial $Y_{\rm p} = 0.247$ and a He-enrichment ratio $\Delta{Y}/\Delta{Z} = 1.31$ \citep{Pietrinferni21}.

\begin{figure}
\centering
\includegraphics[width=0.4\textwidth,keepaspectratio]{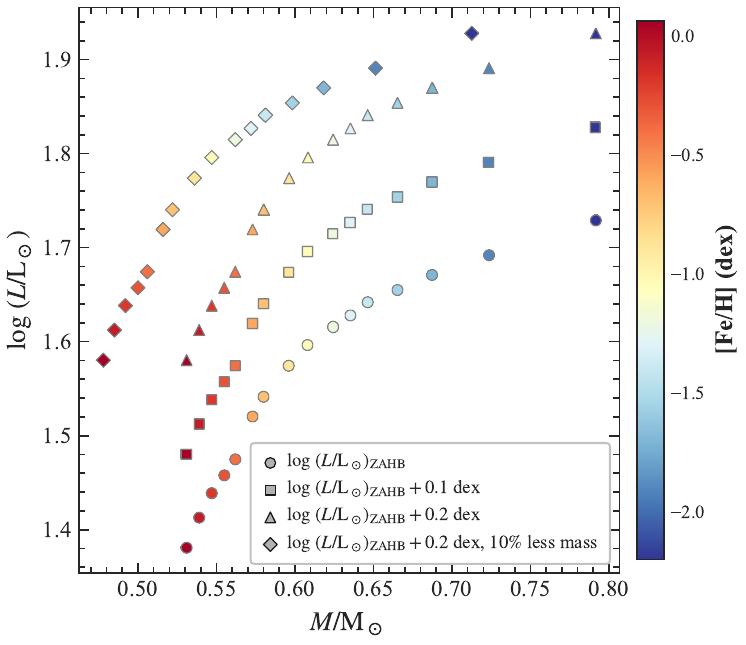}
\caption{Mass–luminosity distribution of the RR~Lyrae model grid. Each chemical composition is represented by four evolutionary configurations: the ZAHB model, two higher-luminosity models at +0.1 and +0.2 dex, and a reduced-mass model at +0.2 dex. Colors indicate metallicity, while different markers denote evolutionary stages.}
\label{fig:MLR}
\end{figure}

For each composition, the BaSTI models provide stellar mass, luminosity, and effective temperature along the ZAHB. As well known from stellar evolution theory, ZAHB luminosities increase with decreasing metallicity, reflecting the larger core mass reached at He ignition. Following \citetalias{marc15}, we account for evolutionary effects and observational spread using two enhanced luminosity levels above the ZAHB ($+0.1$ and $+0.2$ dex), the latter also including 10\% mass loss along the HB. RRLs are expected to occupy the IS temperature range, typically 5000--7400 K, but we considered a wider range ($4000$--$8000$~K) with a 50 K step to ensure full coverage. We extracted the stellar parameters corresponding to this range and used them as input for the linear nonadiabatic (LNA) pulsation calculations. We summerized the input model grid in Table~\ref{tab:zahb_grid}. Fig.~\ref{fig:MLR} shows the mass–luminosity distribution adopted from the BaSTI HB models across the metallicities considered. The full machine-readable table is available at Zenodo: \url{https://doi.org/10.5281/zenodo.20611450}.

\begin{table}
\centering
\small
\setlength{\tabcolsep}{4pt}
\caption{Summary of the adopted input model grid.}
\label{tab:zahb_grid}
\begin{tabular}{lccc}
\hline
Parameter & \multicolumn{3}{c}{Adopted value} \\
\hline
$\mathrm{[Fe/H]}$ range in dex & \multicolumn{3}{c}{$-2.2$ to $+0.06$} \\
Metal fraction $(Z)$ range & \multicolumn{3}{c}{$0.000199$ to $0.032580$} \\
Helium fraction $(Y)$ range & \multicolumn{3}{c}{$0.2472$ to $0.2896$} \\
Mass range ($M/\msun$) & \multicolumn{3}{c}{$0.478$ to $0.791$} \\
$\log(L_{\rm ZAHB}/\lsun)$ range & \multicolumn{3}{c}{$1.381$ to $1.928$} \\
Temperature range & \multicolumn{3}{c}{$4000$ - $8000$~K, in steps of $50$~K} \\
\hline
TC parameters & Set~A & Set~D & \citetalias{kova24} \\
\hline
Mixing-length, $\alpha_{\rm MLT}$ & $1.2$, $1.5$, $1.8$ & $1.2$, $1.5$, $1.8$ & $1.5$ \\
Eddy-viscous dissipation, $\alpha_{m}$ & $0.25$ & $0.70$ & $f(T_{\rm eff})^{m}$ \\
Turbulent source, $\alpha_{s}$ & $\frac{1}{2}\sqrt{\frac{2}{3}}$ & $\frac{1}{2}\sqrt{\frac{2}{3}}$ & $0.32\pm0.06$ \\
Convective flux, $\alpha_{c}$ & $\frac{1}{2}\sqrt{\frac{2}{3}}$ & $\frac{1}{2}\sqrt{\frac{2}{3}}$ & $0.32\pm0.06$ \\
Turbulent dissipation, $\alpha_{d}$ & $\frac{8}{3}\sqrt{\frac{2}{3}}$ & $\frac{8}{3}\sqrt{\frac{2}{3}}$ & $f(T_{\rm eff})^{n}$ \\
Turbulent pressure, $\alpha_{p}$ & $0$ & $\frac{2}{3}$ & $\frac{2}{3}$ \\
Turbulent flux, $\alpha_{t}$ & $0$ & $0.01$ & $0.61\pm0.05$ \\
Radiative cooling, $\gamma_{r}$ & $0$ & $2\sqrt{3}$ & $0$ \\
\hline
\end{tabular}
\tablefoot{\citetalias{kova24} values correspond to the general RRL parameter set. \\
$^{m}$ $\bar\alpha_\nu(T_{\rm eff}) = \dfrac{a}{\exp[b(T_{\rm eff}-c)]+1}+d$, with $a=0.775$, $b=0.0024$, $c=6444$~K, $d=0.01$.\\ $^{n}$ Same functional form, with $a=9.7979$, $b=-0.01$, $c=6950$~K, $d=1.089$.\\
The full model grid is available at \url{https://doi.org/10.5281/zenodo.20611450}.}
\end{table}

Relative to the previous BaSTI $\alpha$-enhanced release, the updated ZAHB models are fainter by ~0.02–0.04 dex in $\log L/\lsun$, with mass shifting by a few percent, the sign depending on metallicity. This offset stems primarily from atomic diffusion in the new calculations, which increases the He-core mass at ignition while depleting envelope He and reducing H-shell burning efficiency; this is partially compensated by updated electron conduction opacities, which decrease the He-core mass at ignition \citep{cass07}. 
These differences arise from revisions in the input physics and chemical compositions, and can produce measurable shifts in the predicted pulsation properties such as IS topology, period-luminosity-metallicity
(PLZ) and period-Wesenheit-metallicity (PWZ) relations, making the updated evolutionary models necessary to refine the theoretical description of RRLs.

\subsection{Pulsation code and numerical framework}
\label{sec:pulsation_model}
The pulsation calculations were carried out with the publicly available \mesarsp\ (version r-23.05.1), which permits both linear and nonlinear modeling of radial pulsations in variables such as classical Cepheids, Type II Cepheids (BL Herculis), RRLs and high-amplitude $\delta$ Scuti stars \citep[e.g.,][]{smol08,paxt19,das21,deka22b,kurb23, kova23,kova24,deka24,deka26,smol26}. Here, we restrict the analysis to LNA models, since for IS boundary locations and mean PL/PW relation coefficients, linear models are adequate \citep{deka24,smol26}.

The input microphysics follows the standard \mesa\ implementation. Thermodynamic quantities are computed using the \mesa\ equation of state, which blends several EOS formalisms across the relevant thermodynamic regimes, including OPAL \citep{roge02}, SCVH \citep{saum95}, FreeEOS \citep{irwi04}, HELM \citep{timm00}, PC \citep{pote10}, and Skye \citep{jerm23}; see \citet{paxt19} for a detailed description. Radiative opacities are taken from the standard \mesa\ tables, based on OPAL opacities \citep{igle96} and supplemented at low temperatures by \citet{ferg05}. These tables were computed using a solar-scaled heavy-element mixture based on \citet{aspl09}. The BaSTI models instead adopt the \citet{caff11} solar mixture, which is not currently available among the standard \mesa\ opacity tables; however, differences between the \citet{aspl09} and \citet{caff11} mixtures are relatively small and are not expected to significantly affect pulsation properties, which primarily depend on total metallicity and $\alpha$-enhancement level. The ZAHB models considered here are $\alpha$-enhanced while the opacity tables are solar-scaled, introducing a minor inconsistency; its impact is expected to be limited, since opacities in the relevant temperature regime depend mainly on global metal content, so residual effects are likely subdominant compared to other uncertainties, particularly those affecting convection and color–temperature relations.

The numerical setup follows our previous implementation in \citet{deka24}, based on the fiducial values in \citet{paxt19}, with the main adopted parameters summarized here. Calculations were performed with \mesarsp, which solves the time-dependent pulsation equations on a Lagrangian mesh partitioned into inner and outer regions by an anchor temperature, $T_{\rm anchor} = 11,000$~K for all models. Our standard setup employed $N_{\rm total} = 200$ mass zones, of which $N_{\rm outer} = 60$ were located above the anchor layer. Where this zoning did not yield a stable initial hydrostatic model, we adopted a reduced mesh with $N_{\rm total} = 150$ and $N_{\rm outer} = 40$. The remaining numerical controls, including the inner boundary temperature $T_{\rm inner} = 2 \times 10^{6}$~K, followed the recommendations of \citet{paxt19}; see also Appendix~B of \citet{deka24} for details. These numerical parameters were kept fixed throughout the full model grid.

\mesarsp\ adopts the time-dependent TC formalism of \citet{Kuhf86}, as implemented by \citet{smol08}. This prescription contains eight dimensionless parameters regulating mixing length, eddy viscosity, turbulent driving and dissipation, convective transport, turbulent pressure, turbulent flux, and radiative losses. \citet{paxt19} provides four combinations of these parameters, labeled A, B, C, and D, differing in the inclusion of radiative cooling, turbulent pressure, and turbulent flux. In Set A, turbulent pressure, turbulent flux, and radiative cooling are neglected, and the static limit of the Kuhfuss formalism becomes locally equivalent to standard mixing-length theory. Set B includes radiative cooling following \citet{wuch98}. Set C includes turbulent pressure and turbulent flux, following \citet{yeck98}. Set D includes all these effects simultaneously. These parameters remain uncalibrated with observations; preliminary attempts to constrain them empirically for RRLs and classical Cepheids exist, but only for relatively small samples, so a generally applicable calibration is still lacking \citep[e.g.,][]{kova23,kova24,deka26}. The systematic exploration of these parameters within \mesarsp\ is important, but a full exploration is computationally very expensive.  We therefore adopt the fiducial convection sets from \citet{paxt19} as a physically motivated starting point, selecting Set~A and Set~D as representative of the simplest and most complex prescriptions, supplemented by the empirically calibrated \citetalias{kova24} set which, to our knowledge, has not previously been investigated for the RRL IS. For set~A and D, we consider three mixing-length values, $\alpha_{\mathrm{MLT}} = 1.2$, $1.5$, and $1.8$, chosen to bracket the range of convective efficiencies reported in the literature depending on pulsation mode, metallicity, and photometric band \citep[e.g.,][]{cris04, Netz23}. However, for \citetalias{kova24} set, we use $\alpha_{\rm MLT}=1.5$ as reported in \citetalias{kova24}.   We caution that the optimal $\alpha_{\mathrm{MLT}}$ depends on the specific convection formulation, so values from different codes or TC schemes are not directly comparable.

\subsection{Determination of pulsational stability}
\label{sec:pulsation_stability}
For each equilibrium model in the input grid, we performed a LNA stability analysis of the first ten radial eigenmodes. Here, however, we restrict the discussion to the two modes relevant for RRL pulsation: the FU (RRab) and FO (RRc) modes. The stability of a given mode was determined from its linear growth rate, obtained from the complex eigenfrequency in the nonadiabatic calculation. Modes with positive growth rates were classified as linearly unstable (excited), those with negative growth rates as linearly stable (damped).
At high effective temperatures, models are stable due to insufficient driving in the partial ionization zones. As effective temperature decreases, the growth rate becomes positive, indicating the onset of pulsational instability; at lower temperatures, increased convective efficiency suppresses pulsations and the growth rate turns negative again. This behavior defines the blue and red IS edges. Including convection in the linear mode of \mesarsp\ enables prediction of both edges.

For each adopted luminosity level, the blue and red IS edges were identified from the bluest and reddest effective temperatures, respectively, at which the mode changes from stable to unstable. The edge temperatures derived at different luminosity levels were then fitted separately with linear relations for the blue and red boundaries. The resulting IS boundaries and their dependence on metallicity and convective parameters are discussed in the following sections.

\section{Results}
\label{sec:results}
\subsection{Instability strip topology in the Hertzsprung–Russell diagram}
The primary aim of this work is to predict the theoretical boundaries of the RRL IS by combining the updated BaSTI HB models with \mesarsp. We first examine the relative behaviour of the IS boundaries within the \mesarsp\ framework, focusing on the effects of convection prescription, mixing-length parameter, and metallicity; comparison with earlier studies is addressed differentially.

\subsubsection{Analytical relations for instability strip boundaries}
\label{sec:IS_relations}

To facilitate comparison with observations and earlier theoretical work, we derive analytical relations for the effective temperature of blue and red IS boundaries as a function of luminosity and metallicity for all model sets, considering two functional forms simultaneously: a linear relation in $\log(L/\lsun)$,

\begin{equation}
\log T_{\rm eff} = \alpha + \beta\log(L/\lsun) + \gamma\log Z ,
\label{eq:IS_linear}
\end{equation}

\noindent and a form adding a quadratic luminosity term,

\begin{equation}
\log T_{\rm eff} = \alpha + \beta\log(L/\lsun)
+ \beta_{2}[\log(L/\lsun)]^{2}
+ \gamma\log Z ,
\label{eq:IS_quad}
\end{equation}

\begin{figure*}
\centering
\includegraphics[width=0.9\textwidth,keepaspectratio]{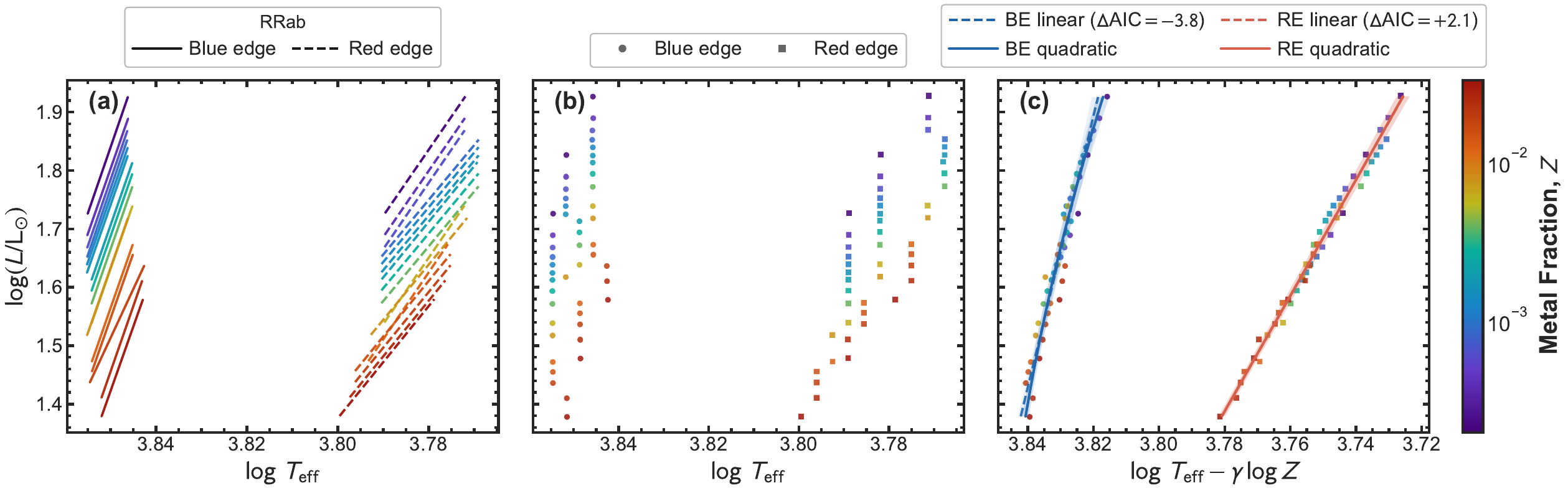}
\caption{
Instability strip boundaries in the Hertzsprung--Russell diagram from the Set~A models with $\alpha_{\rm{MLT}}=1.5$.
Panel~(a) shows the RRab blue and red edges plotted for each metallicity and color-coded by metal fraction $Z$; solid lines denote the blue edge and dashed lines the red edge.
Panel~(b) shows the model points for the same dataset for the blue and red edges in the $\log T_{\mathrm{eff}}$--$\log(L/\lsun)$ plane.
Panel~(c) shows the same points after applying the metallicity correction, $\log T_{\mathrm{eff}} - \gamma \log Z$, together with the fitted linear and quadratic relations for the blue and red edges; the shaded regions represent the corresponding $1\sigma$ uncertainty in the fitted relation.
The common color bar in all panels indicates the metal fraction $Z$.}
\label{fig:IS_RRab}
\end{figure*}
\begin{figure*}
\centering
\includegraphics[width=0.9\textwidth,keepaspectratio]{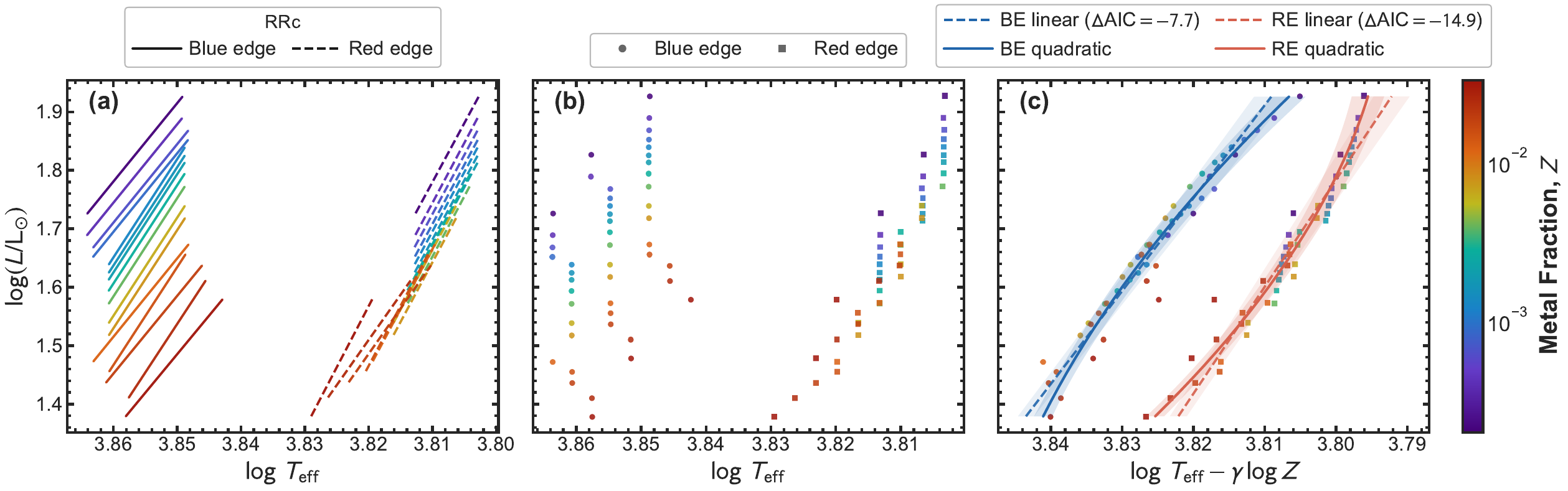}
\caption{
Same as Fig.~\ref{fig:IS_RRab}, but for the RRc.}
\label{fig:IS_RRc}
\end{figure*}

\noindent where $\alpha$ is the zero-point, $\beta$ and $\beta_{2}$ are the linear and quadratic luminosity coefficients, and $\gamma$ is the metallicity coefficient. Both fits are performed simultaneously over all luminosity levels and metallicities available for each model set using ordinary least squares, with coefficients reported in Tables~\ref{tab:IS_relations_FU} and \ref{tab:IS_relations_FO}; the statistically preferred relation for each boundary, selected via the Akaike information criterion and an $F$-test ($\Delta{\rm AIC} = {\rm AIC_{quad}} - {\rm AIC_{lin}} < -2$ and $p<0.05$), is indicated in bold. Representative IS boundaries for Set~A with $\alpha_{\rm MLT}=1.5$ are shown in Figs.~\ref{fig:IS_RRab} and \ref{fig:IS_RRc} for RRab and RRc models, respectively.

We stress an important caveat regarding the validity range of these relations. At fixed metallicity, individual IS boundaries are well-behaved linear functions of luminosity, with the blue edge consistently hotter than the red edge throughout the $\log(L/\lsun)$ range spanned by each metallicity grid (Table~\ref{tab:logL_ranges}). However, when boundaries are expressed as functions of both luminosity and metallicity simultaneously (Eqs.~\ref{eq:IS_quad}), the fitted blue and red edge relations can intersect outside, or near the boundary of, the data range: for the quadratic RRc Set~A, $\alpha_{\rm MLT}=1.8$ fit, the two edges cross near $\log(L/\lsun)\approx1.87$--$1.91$ for $Z\geq0.01208$. Equations~\ref{eq:IS_linear}--\ref{eq:IS_quad} should therefore be applied strictly within the ranges listed in Table~\ref{tab:logL_ranges}; extrapolation beyond these limits may yield unphysical results.

The two modes differ sharply in whether curvature is required. For RRab, the red edge is linear in every configuration, and the blue edge needs a quadratic term only for Set~A (all three $\alpha_{\rm MLT}$) and Set~D at $\alpha_{\rm MLT}=1.8$, always with $\beta_2<0$ (the blue edge cools more slowly at higher $L$). For RRc, by contrast, curvature is the norm rather than the exception: the red edge requires a quadratic term ($\beta_2>0$; the red edge flattens at higher $L$) for every set and $\alpha_{\rm MLT}$ value, and the blue edge is quadratic in all but two of seven configurations. This asymmetry; RRab boundaries close to linear, RRc boundaries intrinsically curved is a recurring pattern that we find again in every photometric plane examined in Sect.~\ref{sec:obs_plane}. The full configuration-by-configuration breakdown is given in Tables~\ref{tab:IS_relations_FU} and \ref{tab:IS_relations_FO}.

The metallicity coefficient $\gamma$ is consistently \emph{negative} at the blue edge across all models and modes: the hot boundary shifts to cooler temperatures at higher $Z$, as expected. At the red edge $\gamma$ is also \emph{negative} in most cases, with one notable exception: the Set~A $\alpha_{\rm MLT}=1.8$ and \citetalias{kova24} RRc red edges show a marginally \emph{positive} $\gamma$ (Table~\ref{tab:IS_relations_FO}). We interpret this weak reversal, which recurs at high convective efficiency, as higher opacity and convective efficiency more readily quenching the FO mode in favour of FU.

\subsubsection{Effect of Mixing-Length Parameter on the Instability Strip}
\label{sec:mixing_length}

TC theory governs the efficiency of convective energy transport in stellar interiors and is a key source of uncertainty in pulsation modelling. Each convective parameter set used here comprises 8 free parameters; we focus exclusively on the mixing-length parameter $\alpha_{\rm MLT}$, keeping all other parameters fixed as defined by Set~A and Set~D. A representative comparison at two metallicities, [Fe/H]~$=-1.55$ and $-0.30$~dex, is shown in Fig.~\ref{fig:mlt_comp}.

For RRab, in agreement with previous theoretical constraints (e.g.\ \citealt{cris04}), IS width decreases systematically with increasing $\alpha_{\rm MLT}$ in both sets (Table~\ref{tab:is_width}). Set~D yields systematically broader strips than Set~A at fixed $\alpha_{\rm MLT}$, with \citetalias{kova24} falling between the two. The metallicity dependence is moderate and prescription-dependent: Set~A narrows monotonically toward higher $Z$, whereas Set~D weakens this trend and even reverses into mild broadening at $\alpha_{\rm MLT}=1.2$; \citetalias{kova24} narrows more shallowly than Set~A.

\begin{table}
\caption{Mean IS width (in $\Delta\log T_{\rm eff}$, with $\Delta T$ in K) as a function of $\alpha_{\rm MLT}$ for RRab and RRc models.}
\label{tab:is_width}
\centering
\begin{tabular}{llccc}
\hline\hline
Type & Set & $\alpha_{\rm MLT}=1.2$ & $\alpha_{\rm MLT}=1.5$ & $\alpha_{\rm MLT}=1.8$ \\
\hline
 & A & 0.085 (1257) & 0.068 (1018) & 0.054 (824) \\
RRab & D & 0.096 (1381) & 0.078 (1138) & 0.061 (899) \\
 & \citetalias{kova24} & -- & 0.074 (1095) & -- \\
\hline
 & A & 0.064 (987)  & 0.043 (675)  & 0.023 (375) \\
RRc  & D & 0.084 (1263) & 0.062 (954)  & 0.037 (564) \\
 & \citetalias{kova24} & -- & 0.052 (810) & -- \\
\hline
\end{tabular}
\tablefoot{$\Delta T$ values (in parentheses) are given in Kelvin. For RRc at $\alpha_{\rm MLT}=1.8$, mean widths are averaged over 14 rather than 16 metallicity points, as the grid does not extend to $Z=0.024270$ and $0.032580$ (see text).}
\end{table}

RRc models show the same qualitative sensitivity to $\alpha_{\rm MLT}$ but occupy a systematically narrower IS, with a markedly stronger dependence on convection prescription than RRab — plausibly because the FO mode is confined to a more external envelope region, where the convective treatment has a proportionally larger effect on the driving/damping balance. Unlike RRab, IS narrowing with increasing metallicity is more clearly established for RRc across all prescriptions, particularly at larger $\alpha_{\rm MLT}$. The RRc grid at $\alpha_{\rm MLT}=1.8$ does not extend to the two highest metallicity points, $Z=0.024270$ and $0.032580$, owing to the absence of stable pulsation solutions there; quoted mean widths for this case are averaged over 14 rather than 16 metallicity points.

The broader strips predicted by Set~D relative to Set~A arise primarily from a displacement of the red edge toward cooler temperatures, with the blue edge comparatively unaffected — consistent with convection's known role in setting the red, rather than blue, boundary. Fig.~\ref{fig:IS_width} shows that this ordering (Set~D $>$ \citetalias{kova24} $>$ Set~A in width, at fixed $\alpha_{\rm MLT}$) and the systematic narrowing with increasing $\alpha_{\rm MLT}$ both hold across the full metallicity range for both modes. Taken together, these results show that IS morphology reflects a coupled dependence on pulsation mode, metallicity, and convective treatment, with the RRc mode particularly more sensitive to convection prescription. This underscores the importance of adopting mode-specific and convection-dependent IS relations when comparing theoretical predictions with observed RRL populations, alongside careful consideration of other uncertainties such as helium abundance and mass loss. The near-closure of the IS at high metallicity has important implications for the predicted pulsation properties and number counts of metal-rich RRL stars (see also \citealt{bono97a,bono97b} for similar results).

\subsubsection{Comparison with \citet{marc15}}
\label{sec:IS_comparison}
We compare our IS boundaries with the theoretical predictions of \citetalias{marc15}, the most comprehensive previous RRL IS study based on nonlinear convective pulsation models ($\alpha_{\rm MLT}=1.5$); whereas our predictions are derived from linear pulsation models. The comparison is shown in Fig.~\ref{fig:M15_comp} at four representative metallicities, for the intermediate value $\alpha_{\rm MLT}=1.5$; the quantitative analysis below extends to all three $\alpha_{\rm MLT}$ values we explored.

The overall IS topology is qualitatively consistent between the two studies. Both predict the IS shifting to cooler temperatures and lower luminosities with increasing metallicity, with comparable luminosity slopes ($|\beta| \approx 0.03$--$0.10$) and metallicity coefficients of similar sign and magnitude — these global trends are therefore robust to the choice of pulsation code.

The primary systematic difference is a zero-point offset in $T_{\rm eff}$. For the blue edge, all our model sets predict cooler temperatures than \citetalias{marc15} by $\approx 10$--$110$~K, with the offset increasing with $\alpha_{\rm MLT}$ in Set~D but decreasing in Set~A. Notably, at $\alpha_{\rm MLT} = 1.5$ the Set~A blue edge ($\Delta\log T_{\rm eff} = -0.0049$, $\approx -34$~K) and the RRc Set~A blue edge ($\Delta\log T_{\rm eff} \approx 0$~K) agree excellently with \citetalias{marc15}, making Set~A at $\alpha_{\rm MLT} = 1.5$ the closest match to the Stellingwerf-code predictions for the blue boundary.

The red edge reveals a more physically significant trend. In Set~A, the offset relative to \citetalias{marc15} increases monotonically with $\alpha_{\rm MLT}$: at $\alpha_{\rm MLT} = 1.2$ the RRab red edge lies $\approx 47$~K cooler than \citetalias{marc15}, whereas at $1.8$ it lies $\approx +172$~K hotter. This sign reversal as convective efficiency increases directly reflects more efficient convective quenching of pulsation driving at higher $\alpha_{\rm MLT}$, which pushes the red edge to higher temperatures. The same reversal occurs in Set~D but at smaller amplitude, with the RRab red edge crossing the \citetalias{marc15} value between $\alpha_{\rm MLT} = 1.5$ and $1.8$. The \citetalias{kova24} prescription, with a mean red edge offset of only $\approx -9$~K, is closest to the \citetalias{marc15} red edge.

For RRc, the red edge is systematically hotter than \citetalias{marc15} across all sets and $\alpha_{\rm MLT}$ values, with offsets ranging from $\approx +26$~K (Set~D, $\alpha_{\rm MLT} = 1.2$) to $\approx +353$~K (Set~A, $\alpha_{\rm MLT} = 1.8$). This offset grows strongly with $\alpha_{\rm MLT}$ in both sets, making the red edge the IS boundary most sensitive to convection prescription. Furthermore, across all metallicities the Set~A red edge offset increases steadily toward higher $Z$ (from $\approx +44$~K at $Z = 0.000199$ to $\approx +113$~K at $Z = 0.015070$), while the blue-edge offset remains broadly constant.

These results, obtained within the framework of linear pulsation analysis,  show that convection prescription introduces systematic IS boundary uncertainties of order $\sim 100$--$350$~K. Building on \citetalias{marc15}, this work extends that framework by exploring three prescriptions at three $\alpha_{\rm MLT}$ values each, providing a quantitative mapping of this uncertainty space over the full metallicity range of observed RRL populations, including metallicities up to approximately solar or slightly super-solar values ($+0.06$~dex) — not covered in \citet{marc15} but covered in \citet{bono97b}. We return to the PLZ/PWZ comparison with \citet{marc21} and \citet{marc22} in the following sections.

To help disentangle whether the $T_{\rm eff}$ offsets reported above originate from the updated BaSTI ZAHB or from the pulsation code itself, we additionally computed non-linear nonadiabatic IS edges with the Stellingwerf code on the same updated ZAHB models adopted for the \mesarsp\ grid, at two representative metallicities ($Z = 0.001$ and $Z = 0.02$, $\alpha_{\rm MLT} = 1.5$), shown in Fig.~\ref{fig:stellingwerf_comparison}. Relative to the previous BaSTI release used by \citetalias{marc15}, the updated ZAHB shifts the mass and luminosity slightly (e.g., at $Z = 0.001$, $M/\msun = 0.64$, $\log L/\lsun = 1.67$ becomes $M = 0.646$, $\log L/\lsun = 1.641$; at $Z = 0.02$, $M/\msun = 0.54$, $\log L/\lsun = 1.49$ becomes $M/\msun = 0.547$, $\log L/\lsun = 1.439$). Comparing the new Stellingwerf models against \citetalias{marc15} isolates the ZAHB update alone (same code): the shift in edge $T{\rm eff}$ is modest ($\lesssim 100$~K), concentrated mainly at the RRc red edge, with the RRab edges and RRc blue edge essentially unchanged within the 
$\pm50$~K uncertainty. Comparing the same models against \textsc{mesa-rsp} instead isolates the effect of the pulsation code (same ZAHB); the level of agreement depends on the convection set. At the RRab edges, Set~D agrees closely with Stellingwerf (differences $\lesssim$50~K at both metallicities), while Set~A predicts a systematically hotter IS at both edges. At the RRc edges, Set~A agrees closely with Stellingwerf at the blue edge, with the red edge slightly bluer by up to $\approx$150~K at the lower metallicity; the two codes agree at $Z=0.02$. Part of any residual offset may stem from the fact that our \textsc{mesa-rsp} edges are derived from linear models, whereas the Stellingwerf edges compared here are nonlinear. This comparison is intended only as a differential check on the code-dependent $T_{\rm eff}$ zero-point at two metallicities; a more extensive recomputation with the Stellingwerf code on the updated BaSTI ZAHB, across a finer metallicity and mass grid, is in preparation.

\subsection{Instability Strip in the Observational Planes}
Bolometric corrections were obtained from the YBC tables using the PHOENIX/BT-Settl library for the selected photometric systems \citep{chen19}. In the adopted PHOENIX grids, the chemical mixture is metallicity-dependent: $\alpha$-enhanced compositions are used at low metallicity, scaled-solar at higher metallicity. For each star, the bolometric correction in a given band is obtained from the tabulated grid as a function of $\log T_{\mathrm{eff}}$, $\log g$, and metallicity, using bilinear interpolation in $\log T_{\mathrm{eff}}$ and $\log g$ within each metallicity grid, then linear interpolation between the two nearest metallicity grids. Absolute magnitudes were computed from
\[
M_{\mathrm{bol}} = M_{\mathrm{bol},\odot} - 2.5 \log \left( \frac{L}{\lsun} \right)
\]
and
\[
M_i = M_{\mathrm{bol}} - BC_i ,
\]
adopting $M_{\mathrm{bol},\odot} = 4.74$ following the IAU 2015 convention \citep{mama15}.

\label{sec:obs_plane}
We derive IS boundary relations in several photometric planes of direct observational relevance. For each plane we adopt the functional form

\begin{equation}
\mathcal{C} = \alpha + \beta\,\mathcal{M} + \beta_2\,\mathcal{M}^2
+ \gamma\,[\mathrm{Fe/H}],
\label{eq:IS_phot}
\end{equation}

\noindent where $\mathcal{C}$ is the color index, $\mathcal{M}$ is the magnitude in the reference band, and $[\mathrm{Fe/H}]$ is the metallicity — the direct observational analogue of Eq.~(\ref{eq:IS_quad}); $\beta_2 \equiv 0$ when the quadratic term is not statistically warranted. The color--magnitude pairs used are $(V-I,I)$ for Johnson--Cousins, $(G_{\rm BP}-G_{\rm RP},G)$ for Gaia, $(g-r,g)$ and $(g-i,g)$ for LSST, and $(J-K_{\rm s},K_{\rm s})$ for 2MASS, with model selection as in Sect.~\ref{sec:IS_relations}. Full coefficient tables for all convection sets and $\alpha_{\rm MLT}$ values are provided at \url{https://doi.org/10.5281/zenodo.20611450}; an example plot is shown in Fig.~\ref{fig:IS_in_cmd}.

Two patterns already established in the $\log T_{\rm eff}$--$\log(L/\lsun)$ plane (Sect.~\ref{sec:IS_relations}) recur across every photometric system examined here: the RRab red edge is linear in essentially every configuration and every plane, while RRc boundaries -- especially the red edge -- require the quadratic form far more often. This confirms the FO curvature is a genuine physical feature rather than an artifact of the temperature--luminosity representation. The metallicity coefficient $\gamma$ is \emph{positive} (more metal-rich RRL are redder at fixed luminosity/magnitude) for essentially every boundary in every plane, with one important, recurring exception discussed below.

In the optical planes ($V-I$, Gaia, LSST $g-i$), $\gamma$ at the RRc red edge turns \emph{negative} for a minority of high-convective-efficiency configurations -- Set~A/D at $\alpha_{\rm MLT}=1.8$ and \citetalias{kova24} are the recurring offenders -- while remaining \emph{positive} or consistent with zero elsewhere (see the relevant coefficient table for exact values per configuration). In the near-infrared $J-K_{\rm s}$ plane, this reversal becomes the \emph{majority} behaviour: $\gamma$ is \emph{negative} for six of seven configurations, with substantially larger amplitude than in any optical plane, indicating the FO red edge becomes progressively bluer with increasing metallicity in the NIR. We interpret both the optical and NIR versions as arising from the same mechanism proposed in Sect.~\ref{sec:mixing_length}: confinement of the FO mode to a more external envelope region, where increased metallicity -- and hence opacity and convective efficiency -- produces earlier quenching. The reversal's greater systematicity and significance in the NIR suggest this effect is best probed there, and its physical origin merits further investigation with non-linear pulsation models.

 A few features are particular to individual systems rather than shared across all of them. The $V-I$ and \textit{Gaia} strips are quantitatively very similar for both modes, as expected given their comparable optical baselines; red-edge $\gamma$ values are somewhat larger in \textit{Gaia} than in $V-I$. The two LSST indices diverge more from each other than from \textit{Gaia} or $V-I$: $(g-i)$'s broader wavelength baseline gives it consistently larger $|\gamma|$ than $(g-r)$ (typically $+0.039$ to $+0.051$ vs.\ $+0.036$ to $+0.044$ at the blue edge, compared to $+0.032$--$0.042$ in \textit{Gaia}), and the RRc blue edge is quadratic in all seven configurations in $(g-r)$ -- the highest incidence of blue-edge curvature found anywhere in this study. The $J-K_{\rm s}$ plane gives the tightest RRab fits of any system studied ($R^2$ values are the highest obtained across all photometric planes), consistent with the known reduced sensitivity of NIR colors to reddening and metallicity \citep{bono01,long86,long90}, and is where the RRc red-edge $\gamma$ reversal, discussed above, is most pronounced.

\subsection{The Kiel Diagram}
Large-scale spectroscopic surveys are now delivering atmospheric
parameters ($T_{\rm eff}$, $\log g$, $[\mathrm{Fe/H}]$) for
thousands of RRL. \citet{wang24} derived parameters for
$\sim$11\,000 RRLs from LAMOST spectra, and the DESI Year~1 catalog
\citep{medi25} has provided the first empirical constraints on the
metallicity-dependent topology of the RRL IS using
phase-corrected effective temperatures for a sample of comparable
size. Upcoming surveys such as 4MOST \citep{dejo19}, WEAVE \citep{dalt12},
and SDSS-V \citep{koll17} will push this further by an order of
magnitude. In this context, the Kiel diagram ($\log T_{\rm eff}$
vs.\ $\log g$) represents one of the most direct planes for
comparing theoretical IS boundaries with spectroscopic observations,
as it is entirely independent of distance, reddening, and photometric
calibration. The boundary relations derived here
(\url{https://doi.org/10.5281/zenodo.20611450}) are therefore directly applicable to
these datasets without any intermediate transformation.

For RRab stars, the linear model is preferred for 12 of 14
configurations, with the quadratic term reaching significance only
for the red edge at the highest $\alpha_{\rm MLT}$ in both Set~A
and Set~D. The metallicity coefficient $\gamma$ is \emph{negative}
throughout, indicating that more metal-rich stars are cooler at
fixed $\log{g}$. 

The RRc boundaries show a markedly different behaviour, with 11 of
14 configurations preferring the quadratic form, a much higher
incidence than for RRab and than is seen in any photometric plane.
The red edge is quadratic for nearly all convection prescriptions,
while the blue edge is quadratic for about half. To our best knowledge, no empirical
constraints on the RRL IS in the Kiel diagram currently exist in the
literature; the relations presented here therefore provide the first
theoretical framework for such a comparison when RRL atmospheric
parameters become available at scale from upcoming surveys.

\subsection{Comparison with \citet{bhar23a}}
\label{sec:is_plz}

Before comparing the predicted PLZ and PWZ relations with observations, we examine how well each convection set reproduces the observed distribution of RRL stars in the $(J-K_s)_0$ vs.\ $M_{K_s}$ CMD, shown in Fig.~\ref{fig:cmd}.

The observed sample is drawn from \citet{bhar23a}, who presented homogeneous $JHK_s$ time-series photometry for 964 RRL variables in 11 Galactic globular clusters spanning $-2.35 \lesssim [\mathrm{Fe/H}] \lesssim -0.44$~dex, obtained with the WIRCam instrument at the Canada-France-Hawaii Telescope (CFHT) and FLAMINGOS-2 at the Gemini South telescope. The sample includes 588 RRab and 345 RRc pulsators, supplemented by 346 Milky Way field RRL stars with $JHK_s$ photometry from 2MASS \citep{skru06} and geometric parallaxes from \textit{Gaia} EDR3 \citep{gaia21, bail21}, restricted to $-2.5 < [\mathrm{Fe/H}] < -0.3$~dex. We corrected the $J$ and $K_s$ magnitudes for reddening following \citet{bhar23a}, adopting the reddening law of \citet{card89} with $R_V = 3.23$ and the total-to-selective absorption ratios in the NIR filters, $A_{J} = 0.94\,E(B-V)$ and $A_{K_s} = 0.39\,E(B-V)$. The $E(B-V)$ values were taken from Table~1 of \citet{bhar23a}.

The blue and red edges of the IS are drawn for two representative metallicities, $[\mathrm{Fe/H}] = -2.2$ and $-0.4$~dex, spanning the full metallicity range of the models. As expected, the red edge is sensitive to $\alpha_\mathrm{MLT}$ while the blue edge remains largely unaffected. This is clearly visible in Fig.~\ref{fig:cmd}: all convection prescriptions produce nearly identical blue edges at $(J-K_s)_0 \approx 0.15$--$0.17$~mag, well within the observed 50 per cent density contour.

For RRab stars (panel~a), Set~D with $\alpha_\mathrm{MLT}=1.2$ produces the widest IS, with its red edge extending well beyond the observed 10 per cent density contour into a region essentially unpopulated by the data, indicating this prescription overestimates the IS width. Notably, the red edge of Set~A with $\alpha_\mathrm{MLT}=1.2$ closely coincides with that of Set~D with $\alpha_\mathrm{MLT}=1.5$, demonstrating a degeneracy between convection set and mixing-length parameter that limits the discriminating power of the CMD comparison alone. At the opposite extreme, Set~A and Set~D with $\alpha_\mathrm{MLT}=1.8$ and the K24 prescription produce narrower strips, with red edges falling near or slightly inside the observed 25 per cent density contour, suggesting a marginal underestimate of the IS width. Among all prescriptions considered, Set~A and Set~D with $\alpha_\mathrm{MLT}=1.5$ provide the most consistent match to the observed red edge locus, coinciding with the 10--25 per cent density contour where the stellar population naturally thins out. A qualitatively similar behaviour is seen for RRc pulsators (panel~b), though the broader density distribution owing to the smaller sample size makes the boundary comparison less discriminating. 


Next, we derive PLZ and PWZ relations in the $JHK_s$ bands from our grid of linear non-adiabatic pulsation models. To ensure consistency with the observed sample of \citet{bhar23a}, we restrict the analysis to $[\mathrm{Fe/H}] < -0.4$~dex and apply period windows of $0.29 < P < 1.00$~d for RRab and $0.20 < P < 0.51$~d for RRc, bracketing the observed period distributions of FU and FO mode pulsators, respectively. The relations are derived separately for RRab, RRc, and a combined RRab+RRc sample; for the latter, RRc periods are converted to the FU mode using $\log P_{\rm RRab} = \log P_{\rm RRc} + 0.127$ \citep{bhar23a, copp15, brag22}.

For each pulsation type and photometric band we fit the linear relation
\begin{equation}
M_\lambda = a + b\log P + c[\mathrm{Fe/H}],
\label{eq:plz}
\end{equation}
using ordinary least squares, where $P$ is the period in days. Formal uncertainties on each coefficient are derived from the regression covariance matrix. In addition to the direct-band relations, three reddening-free Wesenheit magnitudes \citep{mado82} are computed following \citet{bhar16, bhar23a}: $W_{J,H}= H - 1.83(J - H)$, $W_{J,K_s} = K_s - 0.69(J - K_s)$, and $W_{H,K_s} = K_s - 1.92(H - K_s)$.

\begin{figure*}
\centering
\includegraphics[width=0.85\textwidth,keepaspectratio]{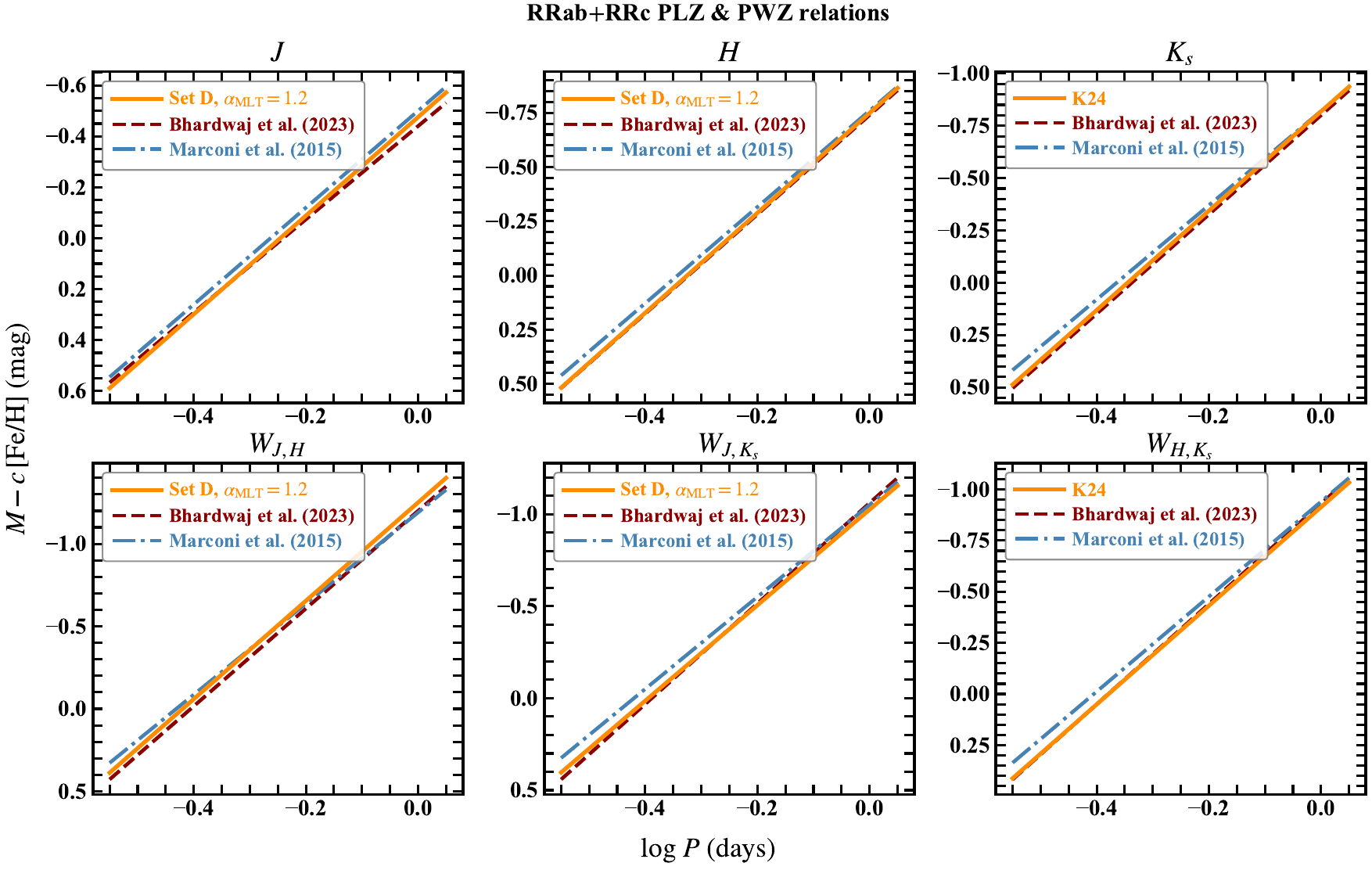}
\caption{The PLZ and PWZ relations in the $JHK_s$ bands
    for the combined RRab$+$RRc sample.
    In each panel the orange solid line shows the best-matching
    theoretical prediction from the present work (convection set
    and $\alpha_\mathrm{MLT}$ labeled), selected by minimizing the
    normalised Euclidean distance in coefficient space to the empirical
    relations of \citet{bhar23a} (dark red dashed).
    The dash-dotted blue line shows the theoretical global
    (RRab$+$RRc) relations of \citet{marc15} for comparison.
    RRc periods are fundamentalized before fitting. The different panels use different y-axis ranges.}
\label{fig:plz_glob}
\end{figure*}

To identify the convection prescription that best reproduces
observations we compare our predicted relations against the empirical
calibration of \citet{bhar23a} for RRab and the combined sample.
For RRc we use the theoretical FO relations of
\citetalias{marc15} as the reference, because a well-constrained
purely empirical RRc calibration in all three $JHK_s$ bands is not
available in \citet{bhar23a}.
The agreement is quantified using a normalised Euclidean distance in
coefficient space,
\begin{equation}
    d = \sqrt{
        \left(\frac{a - a_\mathrm{ref}}{|a_\mathrm{ref}|}\right)^{\!2}
      + \left(\frac{b - b_\mathrm{ref}}{|b_\mathrm{ref}|}\right)^{\!2}
      + \left(\frac{c - c_\mathrm{ref}}{|c_\mathrm{ref}|}\right)^{\!2}
        },
    \label{eq:dist}
\end{equation}
which weights the zero-point $a$, slope $b$, and metallicity
coefficient $c$ equally on a relative scale.
For each band independently we select the convection set that minimizes
$d$; the results are compiled in Tables~\ref{tab:best_fit} and
illustrated in Figs.~\ref{fig:plz_glob}. The PLZ/PWZ relations without any selection criteria on metallicity for all convection is also provided in Table~\ref{tab:plz_combined}.

For the RRab sample, Set~D with $\alpha_\mathrm{MLT} = 1.2$ minimizes
$d$ in four of the six bands ($J$, $H$, $W_{J,H}$, $W_{J,K_s}$),
while the \citetalias{kova24} prescription is preferred in $K_s$ and
$W_{H,K_s}$.
We note, however, that $\alpha_\mathrm{MLT}=1.2$ also produces the
widest IS among the prescriptions considered, which extends the model period baseline and
may influence the regression statistics independently of the physical
accuracy of the red edge.
Among prescriptions with $\alpha_\mathrm{MLT} \geq 1.5$, which are
more consistent with the observed IS width, Set~D with
$\alpha_\mathrm{MLT}=1.5$ provides the most competitive PLZ match
for RRab pulsators.
For the RRc sample, Set~A with $\alpha_\mathrm{MLT}=1.8$ best
reproduces the \citetalias{marc15} FO relations in five of
the six bands.
For the combined RRab$+$RRc sample, Set~D with $\alpha_\mathrm{MLT}=1.2$
is again formally preferred in four bands, subject to the same
caveat noted above, while \citetalias{kova24} is preferred in $K_s$
and $W_{H,K_s}$.

The competitive performance of the \citetalias{kova24} prescription
in the $K_s$ and $W_{H,K_s}$ bands is noteworthy given its empirical
calibration against observations.
However, \citetalias{kova24} provides separate convection parameters for RRab and RRc
pulsators, whereas in this work we adopt their recommended global RRL
parameter set uniformly across both modes for computational
convenience; applying the mode-specific calibrations may further
improve the agreement.

No single prescription provides a uniformly best match across all six
photometric bands, hinting at a possible wavelength dependence of the
preferred convection parameters that may reflect the varying sensitivity
of different bandpasses to the atmospheric temperature structure and
convective flux.
This is an interesting result in its own right and warrants further
investigation with nonlinear models, where full wavelength-dependent
light curves can be predicted and compared directly with multi-band
observations.
We also note that the present analysis is based on LNA
models, which yield IS boundaries and mean magnitudes but carry no
information on pulsation amplitudes or light curve morphology; observables that are sensitive to the convective efficiency
\citep{kova23, kova24, deka26} and would provide important
complementary constraints.
In particular, here the PLZ and PWZ relations are derived using static magnitudes, whereas the observational counterparts are intensity avreaged mean magnitudes, which can be theoretically predicted only in a non-linear approach \citep[see e.g.][and references therein]{marc22}
The implications and limitations of these findings are discussed in Sect.~\ref{sec:discussion}.

\subsection{Period-Wesenheit-Metallicity Relations in \textit{Gaia}}
\label{sec:gaia_plz}
We derived the PWZ relation in the \textit{Gaia} bands and compared it with the empirical relation of \citet{garo22}. We adopted the Wesenheit magnitude formulation of \citet{ripe19},
\begin{align}
    W(G, G_{BP}, G_{RP}) &= G - \lambda\,(G_{BP} - G_{RP}), \notag
\end{align}
with $\lambda = 1.9$, following \citet{ripe19} and \citet{marc21}. We note that the value of $\lambda$ adopted by \citet{garo22} varies slightly with sky position.

\begin{figure*}
\centering
\includegraphics[width=0.75\textwidth,keepaspectratio]{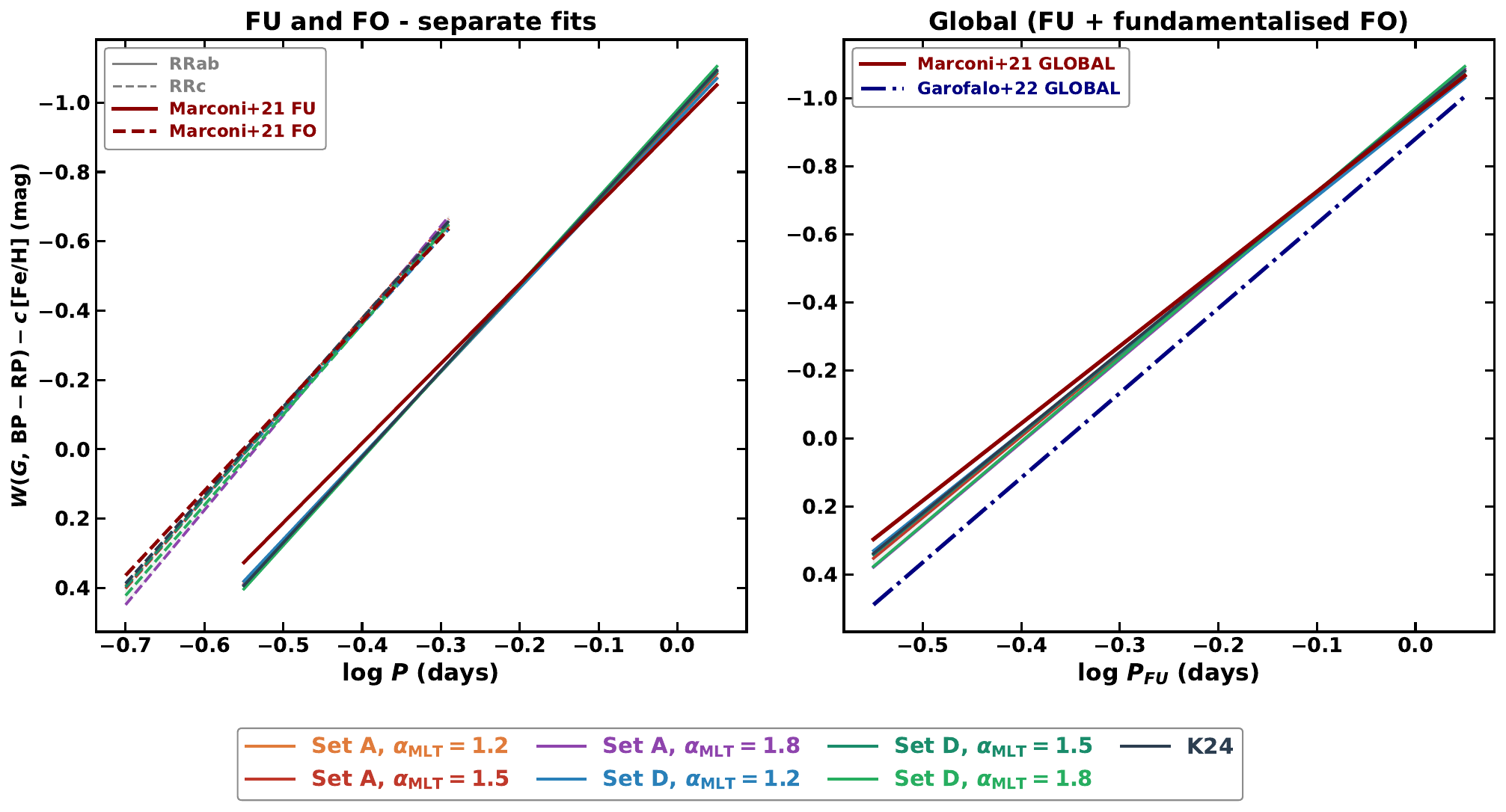}
\caption{\textit{Left:} Metallicity-corrected \textit{Gaia} PWZ relations shown separately for the RRab (solid) and RRc (dashed) pulsators across all convection prescriptions (Set A with $\alpha_{\mathrm{MLT}} = 1.2, 1.5, 1.8$; Set D with $\alpha_{\mathrm{MLT}} = 1.2, 1.5, 1.8$ and K24), compared with the RRab (solid) and RRc (dashed) relations of \citet{marc21}. The different convection prescriptions introduce noticeably more dispersion among the RRc relations than among the FU-mode relations, which remain tightly clustered. \textit{Right:} Global relation (RRab + fundamentalised RRc) for the present models compared with the global relations of \citet{marc21} and \citet{garo22}.}
\label{fig:gaia_pwz_bprp}
\end{figure*}

The derived relations are reported in Table~\ref{tab:pwz_gaia_coeffs}, and their comparison with the theoretical relations of \citet{marc21} and the empirical relations of \citet{garo22} is shown in Fig.~\ref{fig:gaia_pwz_bprp}. Our global relation is in good agreement with \citet{marc21} over the entire period range, while it shows a systematic offset of a few hundredths of a magnitude with respect to the global empirical relation of \citet{garo22}, which increases toward shorter periods. To investigate the origin of this discrepancy, we tested different values of $\lambda$ and examined the period and metallicity ranges spanned by the two calibrations. Adopting the same value of $\lambda$ as \citet{garo22} increases, rather than reduces, the discrepancy. Our model grid spans $\mathrm{[Fe/H]} = -2.2$ to $+0.06$~dex, whereas the \citet{garo22} sample spans $-2.84$ to $+0.07$~dex; we find that this broader and more metal-poor baseline of the empirical calibration is the main driver of the global offset.

When our relation is instead compared directly with the individual \citet{garo22} calibrations for three globular clusters with well-constrained metallicities close to our grid values -- NGC~6121 ($\mathrm{[Fe/H]} = -1.18$), NGC~5904 ($\mathrm{[Fe/H]} = -1.33$), and NGC~3201 ($\mathrm{[Fe/H]} = -1.51$) -- the agreement improves substantially: our models closely reproduce both \citet{marc21} and the cluster-specific relations of \citet{garo22} over the period range covered by the observed stars in each cluster (Fig.~\ref{fig:gaia_pwz}). A small residual offset nonetheless persists in these individual cluster comparisons, which is somewhat unexpected given that the corresponding near-infrared relations show excellent agreement with observations. We further note that the adopted convection prescription introduces substantially more dispersion in the RRc relation than in the RRab relation, for which the different model sets remain tightly clustered (Fig.~\ref{fig:gaia_pwz_bprp}). A detailed characterization of the PL/PW relations is beyond the scope of the present paper; we defer a more thorough investigation to the second paper of this series.

\section{Discussion}
\label{sec:discussion}


The present study is based on LNA pulsation models, computationally efficient and well suited for predicting IS boundaries and mean magnitudes across a wide range of chemical compositions and evolutionary parameters. Stellar evolution inputs play a crucial role in RRL pulsation calculations, since the adopted stellar mass and ZAHB luminosity for a given metallicity directly set the period scale and zero-point of the PLZ/PWZ relations \citep[e.g.,][]{szab04, marc15}. The primary aim of this work is to investigate how updated evolutionary physics, specifically the recent BaSTI HB models \citep{Pietrinferni21}, propagates into the predicted pulsation properties of RRLs, offering a new theoretical framework for their use as distance and metallicity indicators.


The predicted IS edges show the expected behaviour: for both RRab and RRc, the blue and red edges shift redward with increasing metallicity, regardless of convection parameters, consistent with earlier predictions (\citet{bono94}, \citetalias{marc15}). The magnitude of this shift, however, is highly sensitive to the convection prescription, particularly at the red edge.

An interesting behaviour is seen for the RRc red edge: rather than shifting monotonically redward with metallicity, it initially moves redward before turning bluer at higher metallicities. This non-monotonic trend is a distinctive feature of FO pulsation, explainable by the confinement of FO pulsation to a more external envelope region than the FU mode, though it warrants further investigation with nonlinear models. Furthermore, $\alpha_\mathrm{MLT}=1.8$ appears too restrictive for RRc pulsators in the metal-rich regime: for $[\mathrm{Fe/H}] \gtrsim -0.3$~dex, no stable RRc models are found, suggesting this convective efficiency suppresses FO pulsation entirely at higher metallicities.

We also note that \citet{reye24} recently compiled a catalogue of 2824 RRL stars in 115 Galactic globular clusters using \textit{Gaia} DR3 astrometry and photometry, comparing observed IS boundaries with \mesarsp\ models at fixed convection parameters while varying helium abundance. They found $Y = 0.290$ broadly reproduces the observed IS for both modes, but a higher value ($Y = 0.357$) is required for RRc stars, while a lower value ($Y = 0.220$) is favoured for RRab, suggesting mode-dependent sensitivity of the inferred IS topology to helium content. Here, however, we show that variations in metallicity and convection prescription, together with helium abundance determined self-consistently via the enrichment ratio $\Delta Y/\Delta Z$ \citep{Pietrinferni21}, can produce comparable shifts in the predicted IS boundaries. We emphasise that convection in 1D nonlinear pulsation models remains necessarily parametrized, and multidimensional radiation-hydrodynamical calculations would ultimately be required to calibrate these prescriptions from first principles. Nevertheless, our results demonstrate that the inferred IS topology is not driven by helium abundance alone: helium content, metal fraction, and convective efficiency are coupled ingredients whose combined effect must be considered when interpreting observed IS morphologies. A self-consistent exploration of this multidimensional parameter space, ideally informed by future 2D/3D hydrodynamical constraints on convection, represents a compelling direction for future work.

The formal preference for $\alpha_\mathrm{MLT}=1.2$ in the NIR PLZ/PWZ metric most plausibly reflects the wider IS produced by less efficient convection: a broader strip increases the model sample size and extends the period baseline, mechanically improving the regression slope without implying more accurate red-edge predictions. The CMD comparison in Sect.~\ref{sec:is_plz} offers an independent check, showing that $\alpha_\mathrm{MLT}=1.2$ red edges extend into regions of the observational plane essentially unpopulated by the data. This underscores the value of combining multiple diagnostics — IS boundaries, PLZ/PWZ coefficients, and light curve morphology — rather than relying on any single metric.

The \citetalias{kova24} prescription gives the best match in the $K_s$ and $W_{H,K_s}$ bands for both RRab and the combined sample, notable given this set is empirically calibrated against observed RRL light curves. Two caveats apply, however. First, \citetalias{kova24} provides separate parameters for RRab and RRc, whereas we adopt their recommended global RRL set uniformly across both modes for computational convenience; the mode-specific calibrations may improve agreement further, particularly for RRc. Second, the band-dependent preference for different prescriptions in the PLZ/PWZ comparison hints at a possible wavelength dependence of the optimal convection parameters, which current 1D treatments cannot simultaneously accommodate. Exploring this with a finer grid of TC parameters is an important avenue for future work.

For the \textit{Gaia} bands, our predicted PWZ relations agree well with \citet{marc21} for the RRab, RRc, and global samples alike. The choice of convection prescription introduces substantially more scatter in the RRc relations than in RRab, with RRab coefficients remaining tightly clustered across sets. This RRc-driven scatter propagates into the global relation, which combines fundamentalised RRc periods with the RRab sample. Compared to the empirical Milky Way field RRL PWZ relation, however, our models show a significantly larger discrepancy, which we suspect primarily reflects a mismatch in metallicity coverage between theoretical and empirical samples rather than an intrinsic shortcoming of the models. In support of this, the relation from Set~A ($\alpha_{\rm MLT}=1.5$) gives the closest match to the empirical PWZ relations of the three Galactic globular clusters considered (NGC~6121, NGC~5904, and NGC~3201), whose metallicity ranges better match our grid.

We also note that, although \citetalias{kova24} is empirically calibrated against observed RRL pulsation properties, it is not adopted as a reference set here. Their calibration sample spans a more limited range of periods and metallicities than current RRL catalogues and the model grid considered here. Consequently, the prescription is well motivated within its calibration parameter space, but cannot be assumed to uniquely describe convection across the full range of RRL properties. Moreover, as shown in Sect.~\ref{sec:results}, the predicted IS boundaries and PLZ/PWZ relations depend jointly on metallicity and convection prescription. For this reason, this work explores a broad range of convection parameters rather than a single calibrated configuration.


\section{Summary and Conclusions}
\label{sec:summary}
We presented linear pulsation model predictions for RRL stars based on updated BaSTI HB evolutionary tracks and the \mesarsp\ module, focusing on IS topology for both FU and FO modes across seven convection sets. The IS boundaries were transformed into multiple observational photometric systems for direct comparison with current and forthcoming surveys, including \textit{Gaia}, OGLE, 2MASS, LSST, and the Kiel diagram. PLZ and PWZ relations were derived in the near-infrared $JHK_{\rm s}$ bands and in the \textit{Gaia} bands. The NIR relations were compared against the empirical relations of \citet{bhar23a} for RRab and combined samples, and \citetalias{marc15} for RRc stars, while the \textit{Gaia} PWZ relation was compared against \citet{marc21} and \citet{garo22}. The main findings are summarised as follows:

\begin{itemize}

\item \textbf{IS topology and analytical relations:} We derived analytical relations for the blue and red IS boundaries as a function of luminosity and metallicity, in linear and quadratic forms, for both modes across all seven sets. The quadratic term is warranted far more often for RRc than RRab, particularly at the red edge, indicating intrinsic curvature of the FO IS with luminosity across all planes examined. The metallicity coefficient $\gamma$ is consistently \emph{negative} at the blue edge for both modes, confirming the hot boundary shifts to cooler temperatures at higher $Z$.

\item \textbf{Effect of convection prescription and $\alpha_{\rm MLT}$:} IS width decreases systematically with increasing $\alpha_{\rm MLT}$ for both modes and all sets, consistent with previous nonlinear analyses. Set~D produces broader strips than Set~A at fixed $\alpha_{\rm MLT}$, driven by a redward displacement of the red edge; \citetalias{kova24} falls consistently between the two. RRc is substantially more sensitive to convection prescription than RRab, with width differences up to ${\sim}280$~K between Set~A and Set~D at $\alpha_{\rm MLT} = 1.5$.

\item \textbf{Comparison with \citetalias{marc15}:} The global IS topology is qualitatively consistent with \citetalias{marc15}, with comparable luminosity and metallicity coefficients. The main difference is a $T{\rm eff}$ zero-point offset of ${\sim}100$--$350$~K in the red edge, depending on mode, convection set, and $\alpha{\rm MLT}$; the blue edge offset is small in all cases. At $\alpha_{\rm MLT} = 1.5$, Set~A best matches the \citetalias{marc15} RRc red edge, while Set~D best matches the RRab red edge.

\item \textbf{IS boundaries in observational planes:} We provide IS boundary relations in Johnson-Cousins $(V-I, I)$, \textit{Gaia} $(G_{\rm BP}-G_{\rm RP}, G)$, LSST $(g-r, g)$ and $(g-i, g)$, 2MASS $(J-K{\rm s}, K{\rm s})$, and Kiel ($\log T_{\rm eff}$--$\log g$) planes. Patterns from the $L$--$T_{\rm eff}$ plane — quadratic RRc boundaries and the \emph{negative} metallicity dependence of the RRc red edge at high convective efficiency recur across all systems. However, the NIR reversal of the RRc red-edge metallicity coefficient is more systematic and significant than in the optical.

\item \textbf{CMD comparison:} In $(J-K_s)_0$ vs.\ $M_{K_s}$, Set~D at $\alpha_{\rm MLT} = 1.2$ overestimates IS width, while Set~A and Set~D at $\alpha_{\rm MLT} = 1.8$ and \citetalias{kova24} marginally underestimate it. Set~A and Set~D at $\alpha_{\rm MLT} = 1.5$ best match the observed red-edge locus for both modes.

\item \textbf{PLZ and PWZ relations in NIR:} Formally, Set~D at $\alpha_{\rm MLT} = 1.2$ minimizes the normalised distance to \citet{bhar23a} in most bands; however, it also produces the widest IS, which may artificially extend the model period baseline. Among prescriptions consistent with the observed IS width ($\alpha_{\rm MLT} \geq 1.5$), Set~D at $\alpha_{\rm MLT} = 1.5$ is most competitive for RRab. \citetalias{kova24} performs competitively in $K_s$ and $W_{H,K_s}$. No single prescription matches best across all bands, suggesting a wavelength dependence of the preferred parameters.

\item \textbf{PWZ relations in \textit{Gaia}:} The global \textit{Gaia} PWZ fit agrees well with \citet{marc21} across the full period range, but shows a systematic offset from the empirical relation of \citet{garo22}, attributed to the latter's broader, more metal-poor calibration baseline. Agreement improves substantially against individual globular clusters with metallicities close to our grid, with Set~A at $\alpha_{\rm MLT} = 1.5$ most consistent. As in the NIR, convection prescription introduces greater dispersion in the RRc relation than the RRab, where all prescriptions remain tightly clustered.

\end{itemize}

This paper is the first in a series on a comprehensive theoretical characterization of RRL pulsation properties using updated BaSTI evolutionary inputs and \mesarsp. The grid extends to near-solar and mildly super-solar metallicities ($[\rm{Fe/H}]$ up to $+0.06$~dex), extending the theoretical IS framework of \citetalias{marc15} for comparison with modern surveys sampling RRL stars across a broad metallicity baseline. A natural next step is a complementary grid of nonlinear convective models using the same BaSTI inputs, enabling direct comparison with observed pulsation amplitudes and light curve morphology across the full metallicity range, and providing tighter constraints on the preferred convection prescription and a more definitive assessment of the wavelength-dependent behaviour identified here. Ongoing and forthcoming surveys, including \textit{Gaia} DR4, LSST, Roman, and 4MOST, will substantially increase the number of RRL stars with homogeneous multi-band photometry and spectroscopic metallicities, providing the observational leverage needed to fully exploit this framework.

\section*{Data availability}
The input model grid, \mesarsp\ inlist and the IS boundary
coefficient tables discussed in Sects.~\ref{sec:data}--\ref{sec:results}
are available in machine-readable form at Zenodo,
\url{https://doi.org/10.5281/zenodo.20611450} \citep{deka2026_zenodo}.

\section*{Acknowledgements}
We thank the anonymous referee for the valuable comments and suggestions which have significantly improved the manuscript. MD acknowledges funding from the INAF 2023 Large Grant MOVIE (PI: Marcella Marconi) and project PRIN MUR 2022 (code 2022ARWP9C) ``Early Formation and Evolution of Bulge and HalO (EFEBHO)", PI: Marcella Marconi,  funded by European Union – Next Generation EU. MD also acknowledges the use of the High-Performance Computing facility Pegasus at IUCAA, India. To compute the models, we have used \textsc{\mesa}~r23.05.1  \citep{paxt10,paxt13,paxt15,paxt18,paxt19}. We acknowledge the use of AI language tools for text refinement. AB thanks the funding from the Anusandhan National Research Foundation (ANRF) under the Prime Minister Early Career Research Grant scheme (ANRF/ECRG/2024/000675/PMS). GDS acknowledges funding from \textit{Gaia} DPAC through INAF and ASI (2025-10-HH.0; PI: M. G. Lattanzi) and from INFN, Naples Section, through the QGSKY and Moonlight2 initiatives. This research was supported by the International Space Science Institute (ISSI) in Bern/Beijing through ISSI/ISSI-BJ International Team project ID \#24-603 - “EXPANDING Universe” (EXploiting Precision AstroNomical Distance INdicators in the Gaia Universe).

\bibliographystyle{aa}
\bibliography{reference}

\begin{appendix}
\onecolumn
\section{Model grid}
Table~\ref{tab:logL_ranges} provides the luminosity/magnitude ranges over which the IS relations derived in this work are applicable.

\begin{table*}
\centering
\small
\caption{Lower and upper boundaries of $\log(L/\lsun)$, photometric
magnitudes, and surface gravity for each metallicity $Z$ considered in
this work.}
\label{tab:logL_ranges}
\setlength{\tabcolsep}{3.0pt}
\renewcommand{\arraystretch}{1.2}
\begin{tabular}{c c cc cc cc cc cc cc}
\hline\hline
[Fe/H] &$Z$
  & $\log(L/L_{\odot})_{\rm min}$ & $\log(L/L_{\odot})_{\rm max}$
  & $I_{\rm min}$ & $I_{\rm max}$
  & $G_{\rm min}$ & $G_{\rm max}$
  & $g_{\rm LSST,min}$ & $g_{\rm LSST,max}$
  & $K_{s,{\rm min}}$ & $K_{s,{\rm max}}$
  & $(\log g)_{\rm min}$ & $(\log g)_{\rm max}$ \\
\hline
$-2.20$ & $0.000199$ & $1.7257$ & $1.9271$ & $-0.6675$ & $+0.1293$ & $-0.1269$ & $+0.4214$ & $-0.0093$ & $+0.7861$ & $-1.6349$ & $-0.2578$ & $2.2805$ & $3.0136$ \\
$-1.90$ &$0.000397$ & $1.6887$ & $1.8901$ & $-0.5832$ & $+0.2180$ & $-0.0418$ & $+0.5060$ & $+0.0758$ & $+0.8714$ & $-1.5453$ & $-0.1647$ & $2.2782$ & $3.0115$ \\
$-1.70$ & $0.000628$ & $1.6677$ & $1.8691$ & $-0.5369$ & $+0.2662$ & $+0.0048$ & $+0.5507$ & $+0.1221$ & $+0.9162$ & $-1.4958$ & $-0.1134$ & $2.2765$ & $3.0103$ \\
$-1.55$ &$0.000886$ & $1.6517$ & $1.8531$ & $-0.5015$ & $+0.3029$ & $+0.0413$ & $+0.5869$ & $+0.1643$ & $+0.9541$ & $-1.4580$ & $-0.0743$ & $2.2782$ & $3.0121$ \\
$-1.40$ & $0.001250$ & $1.6387$ & $1.8401$ & $-0.4827$ & $+0.3294$ & $+0.0688$ & $+0.6134$ & $+0.1925$ & $+0.9821$ & $-1.4624$ & $-0.0469$ & $2.2595$ & $3.0125$ \\
$-1.30$ & $0.001572$ & $1.6247$ & $1.8261$ & $-0.4513$ & $+0.3594$ & $+0.1002$ & $+0.6441$ & $+0.2252$ & $+1.0144$ & $-1.4292$ & $-0.0168$ & $2.2669$ & $3.0191$ \\
$-1.20$ &$0.001975$ & $1.6127$ & $1.8143$ & $-0.4249$ & $+0.3845$ & $+0.1264$ & $+0.6697$ & $+0.2527$ & $+1.0416$ & $-1.4005$ & $+0.0085$ & $2.2717$ & $3.0235$ \\
$-1.05$ &$0.002785$ & $1.5937$ & $1.7951$ & $-0.3827$ & $+0.4247$ & $+0.1678$ & $+0.7107$ & $+0.2967$ & $+1.0847$ & $-1.3543$ & $+0.0489$ & $2.2800$ & $3.0312$ \\
$-0.90$ &$0.003920$ & $1.5718$ & $1.7731$ & $-0.3408$ & $+0.4725$ & $+0.2164$ & $+0.7639$ & $+0.3398$ & $+1.1549$ & $-1.3365$ & $+0.0992$ & $2.2751$ & $3.0449$ \\
$-0.70$ &$0.006170$ & $1.5387$ & $1.7402$ & $-0.2732$ & $+0.5554$ & $+0.2895$ & $+0.8371$ & $+0.4117$ & $+1.2325$ & $-1.2903$ & $+0.1973$ & $2.2801$ & $3.0775$ \\
$-0.60$ & $0.007730$ & $1.5177$ & $1.7192$ & $-0.2242$ & $+0.6031$ & $+0.3363$ & $+0.8867$ & $+0.4644$ & $+1.2854$ & $-1.2369$ & $+0.2470$ & $2.2985$ & $3.0933$ \\
$-0.40$ & $0.012080$ & $1.4727$ & $1.6742$ & $-0.1256$ & $+0.7065$ & $+0.4410$ & $+0.9954$ & $+0.5677$ & $+1.4148$ & $-1.1564$ & $+0.3573$ & $2.3210$ & $3.1300$ \\
$-0.30$ & $0.015070$ & $1.4557$ & $1.6572$ & $-0.0879$ & $+0.7445$ & $+0.4774$ & $+1.0331$ & $+0.6066$ & $+1.4568$ & $-1.1119$ & $+0.4000$ & $2.3358$ & $3.1415$ \\
$-0.20$ & $0.018750$ & $1.4367$ & $1.6382$ & $-0.0500$ & $+0.7873$ & $+0.5200$ & $+1.0811$ & $+0.6510$ & $+1.5228$ & $-1.0953$ & $+0.4476$ & $2.3324$ & $3.1543$ \\
$-0.08$ & $0.024270$ & $1.4107$ & $1.6123$ & $+0.0047$ & $+0.8464$ & $+0.5786$ & $+1.1405$ & $+0.7091$ & $+1.5875$ & $-1.0617$ & $+0.5122$ & $2.3369$ & $3.1739$ \\
$+0.06$ & $0.032580$ & $1.3787$ & $1.5803$ & $+0.0744$ & $+0.9177$ & $+0.6513$ & $+1.2144$ & $+0.7785$ & $+1.6698$ & $-1.0136$ & $+0.5887$ & $2.3480$ & $3.1995$ \\
\hline
\end{tabular}
\tablefoot{The reported ranges are the extremes over all
convection prescriptions, $\alpha_{\rm MLT}$ values, and both RRab and
RRc modes at each $Z$. The $\log L$ ranges follow the model grid
definition; the RRc grids at
$\alpha_{\rm mlt}=1.8$ extend only to $Z=0.018750$. Magnitudes are in the Vega system ($I$, $G$, $K_s$) and AB system (LSST $g$); all values are absolute. Surface gravity $\log g$ is in cgs units ($\log[\mathrm{cm\,s^{-2}}]$).}
\end{table*}

\section{IS topology}
\begin{figure}
\centering
\includegraphics[width=0.7\textwidth,keepaspectratio]{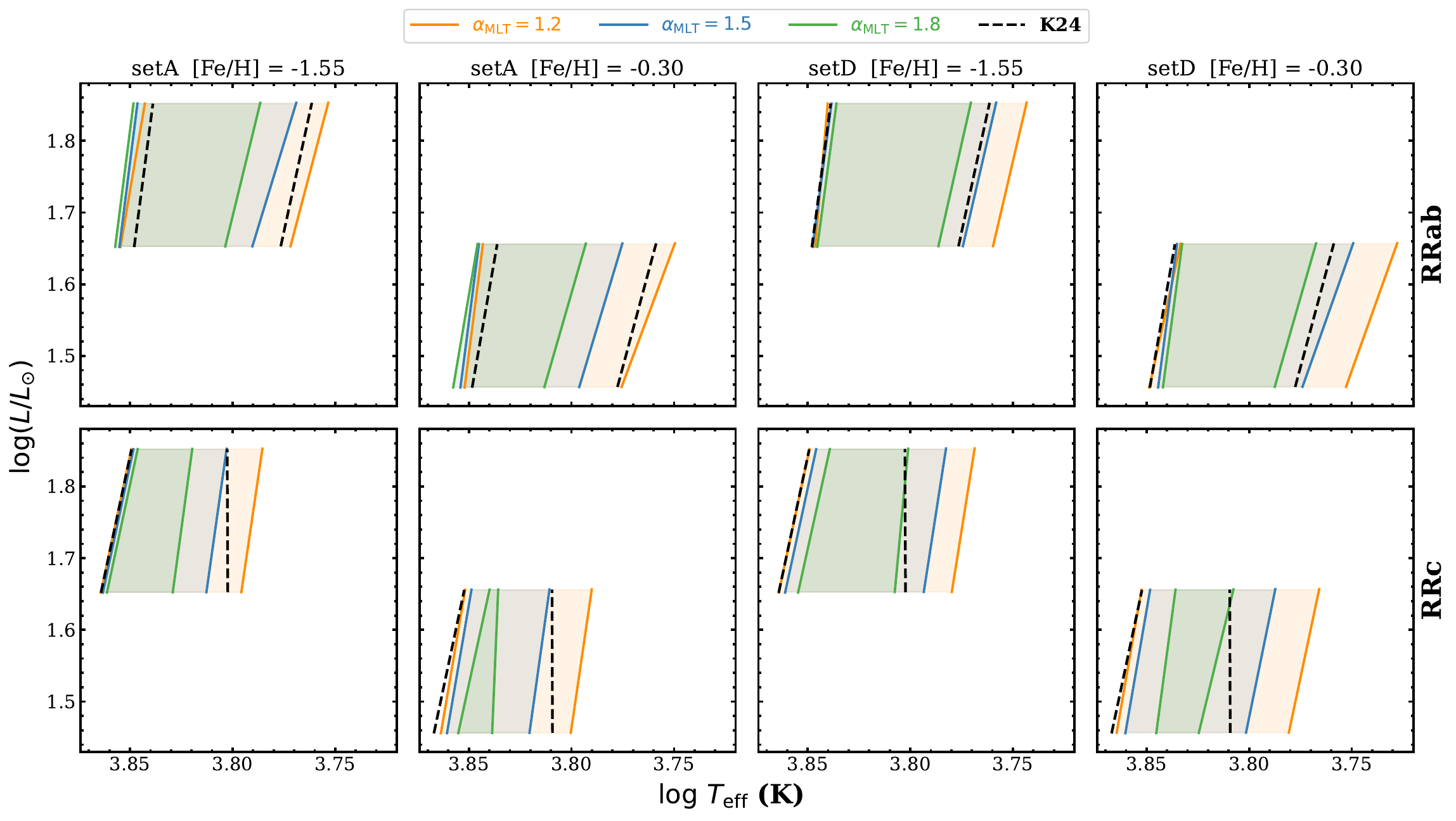}
\caption{IS edges for RRLs at two representative metallicities, $\rm[Fe/H] = -1.55$ (left panels) and $\rm[Fe/H] = -0.30$ (right panels), computed with two sets of convective parameters: Set~A (columns 1--2) and Set~D (columns 3--4). The top and bottom rows show the RRab and RRc IS edges, respectively. Within each panel, the blue (hot) and red (cool) edges are shown for three values of the mixing-length parameter, $\alpha_{\rm MLT} = 1.2$ (red), $1.5$ (blue), and $1.8$ (green), with the shaded region indicating the strip width. The black dashed lines show the IS edges for the \citetalias{kova24} parameter set overplotted for comparison.}
\label{fig:mlt_comp}
\end{figure}

\begin{figure}
\centering
\includegraphics[width=0.8\textwidth,keepaspectratio]{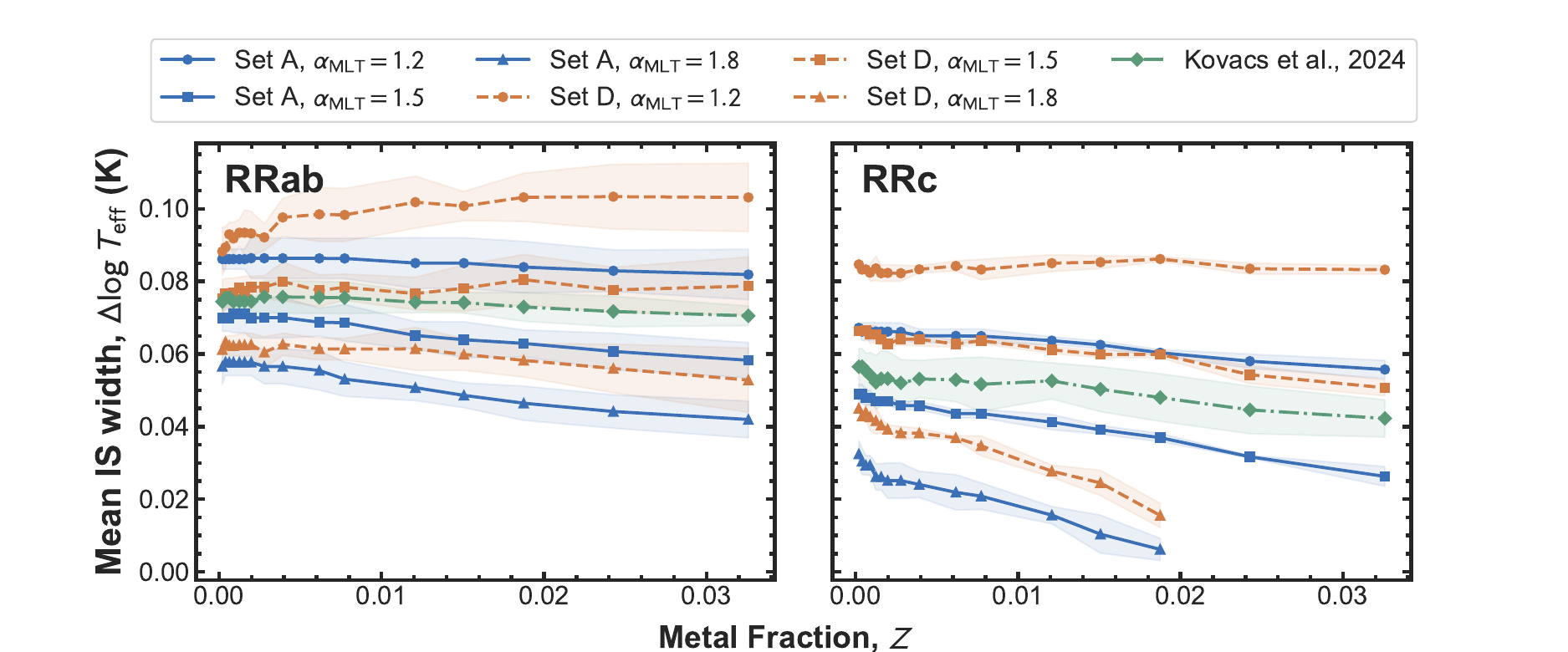}
\caption{
Mean instability strip width as a function of metallicity for RRab and RRc models under different convective assumptions. The left and right panels show the results for RRab and RRc models, respectively. In each panel, width predicted by set~A, set~D, and the \citetalias{kova24} parameters sets  are shown separately, while different symbols identify the adopted values of the mixing-length parameter, $\alpha_{\rm MLT}=1.2$, 1.5, and 1.8. The plotted curves trace the mean width at each metallicity, and the shaded regions indicate the corresponding dispersion within each metallicity bin.
}
\label{fig:IS_width}
\end{figure}

\begin{figure}
\centering
\includegraphics[width=0.8\textwidth,keepaspectratio]{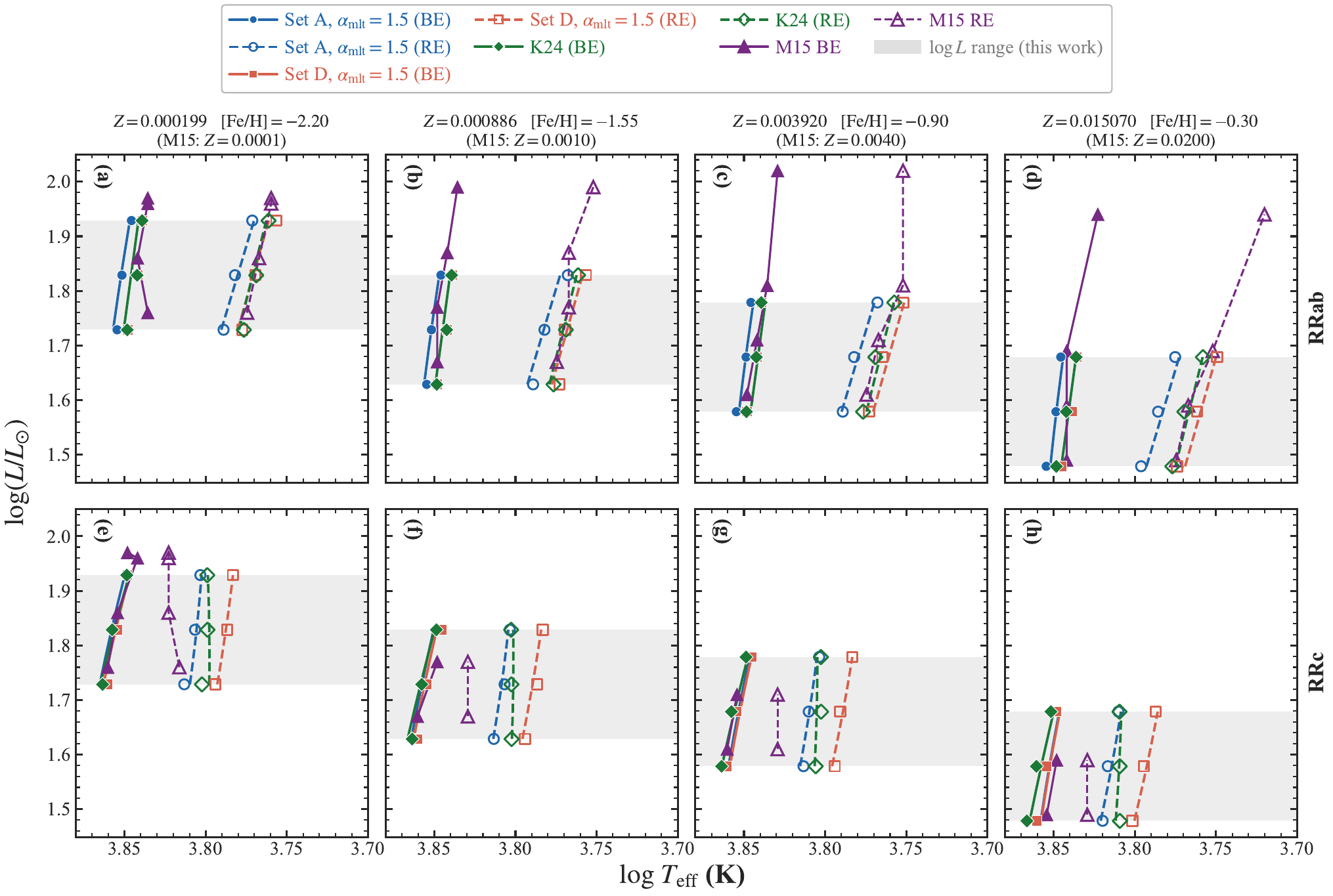}
\caption{Comparison of the theoretical IS boundaries for four representative metallicities (columns, left to right: 
[Fe/H]=-2.20, -1.55, -0.90 and -0.30, respectively) spanning the range from very metal-poor halo populations to the metal-rich Galactic bulge. The upper and lower rows show results for RRab and RRc stars, respectively.}
\label{fig:M15_comp}
\end{figure}

\begin{figure*}
\centering
\includegraphics[width=0.8\textwidth,keepaspectratio]{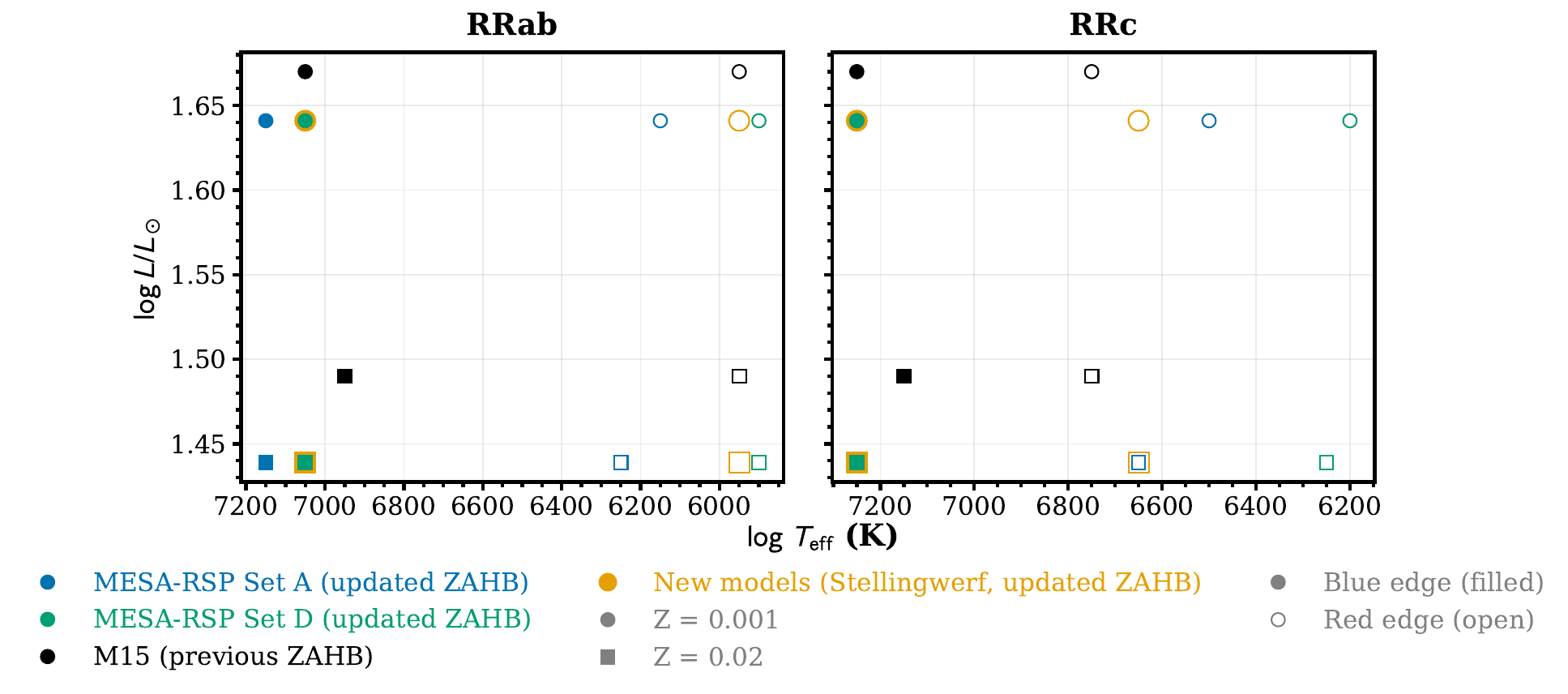}
\caption{
Comparison of instability strip edge temperatures for RRab
(left) and RRc (right) pulsators at two metallicities,
$Z = 0.001$ and $Z = 0.02$ ($\alpha_{\rm MLT} = 1.5$), obtained with
three model sets: \citetalias{marc15} (non-linear Stellingwerf model, previous
BaSTI ZAHB), \textsc{mesa-rsp} (linear, updated BaSTI ZAHB), and new
non-linear Stellingwerf models computed on the same updated BaSTI
ZAHB as \textsc{mesa-rsp}. Filled and open symbols mark the blue and
red edges, respectively; circles and squares denote
$Z = 0.001$ and $Z = 0.02$.
}
\label{fig:stellingwerf_comparison}
\end{figure*}

\begin{figure*}
\centering
\includegraphics[width=0.8\textwidth,keepaspectratio]{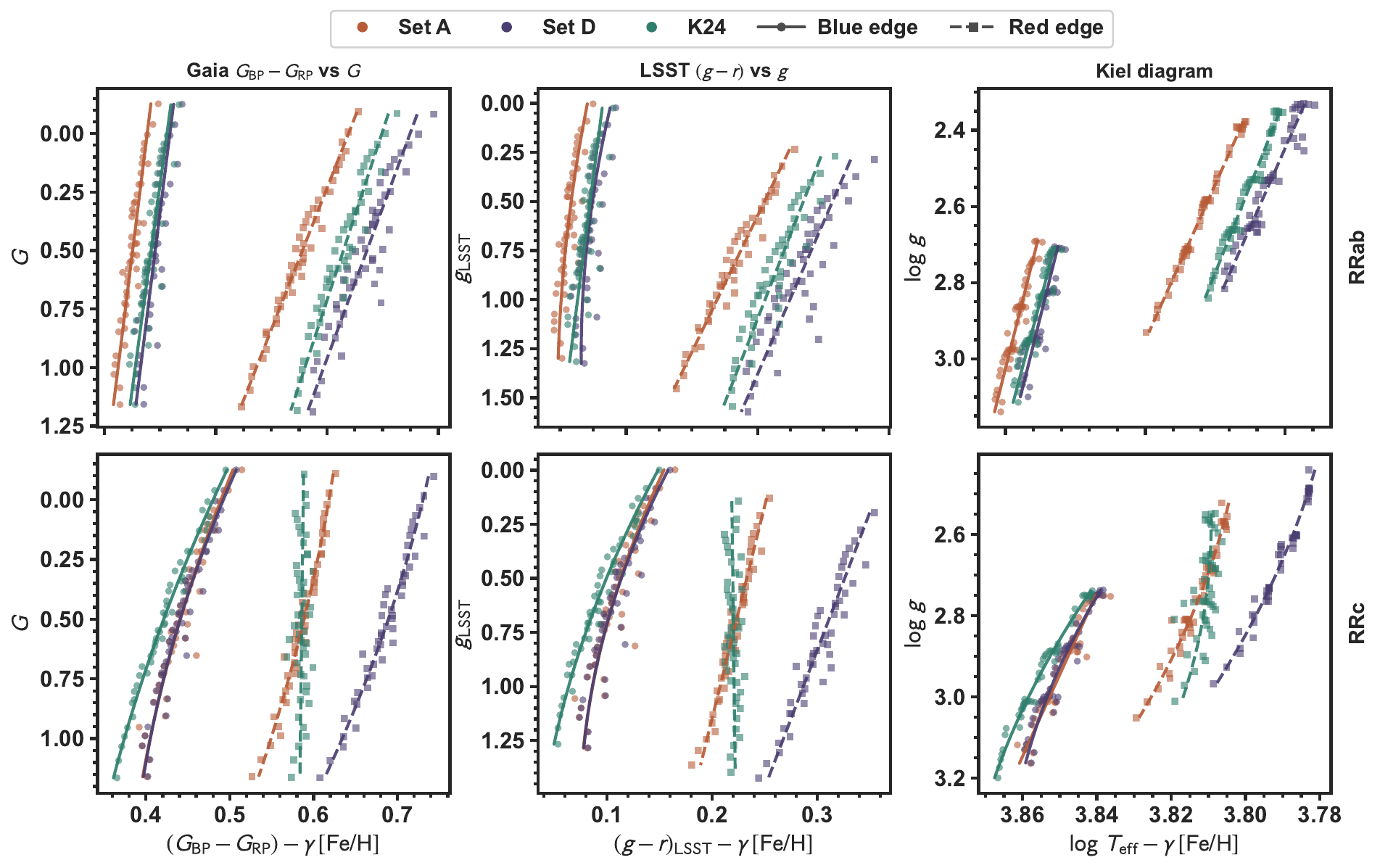}
\caption{Theoretical IS boundaries with convection sets: set~A, set~D, and \citetalias{kova24} ($\alpha_{\rm MLT}=1.5$), shown in the \textit{Gaia}, LSST, and Kiel diagram planes. The upper and lower rows show results for RRab and RRc stars, respectively.}
\label{fig:IS_in_cmd}
\end{figure*}

\begin{figure}
\centering
\includegraphics[width=0.8\textwidth,keepaspectratio]{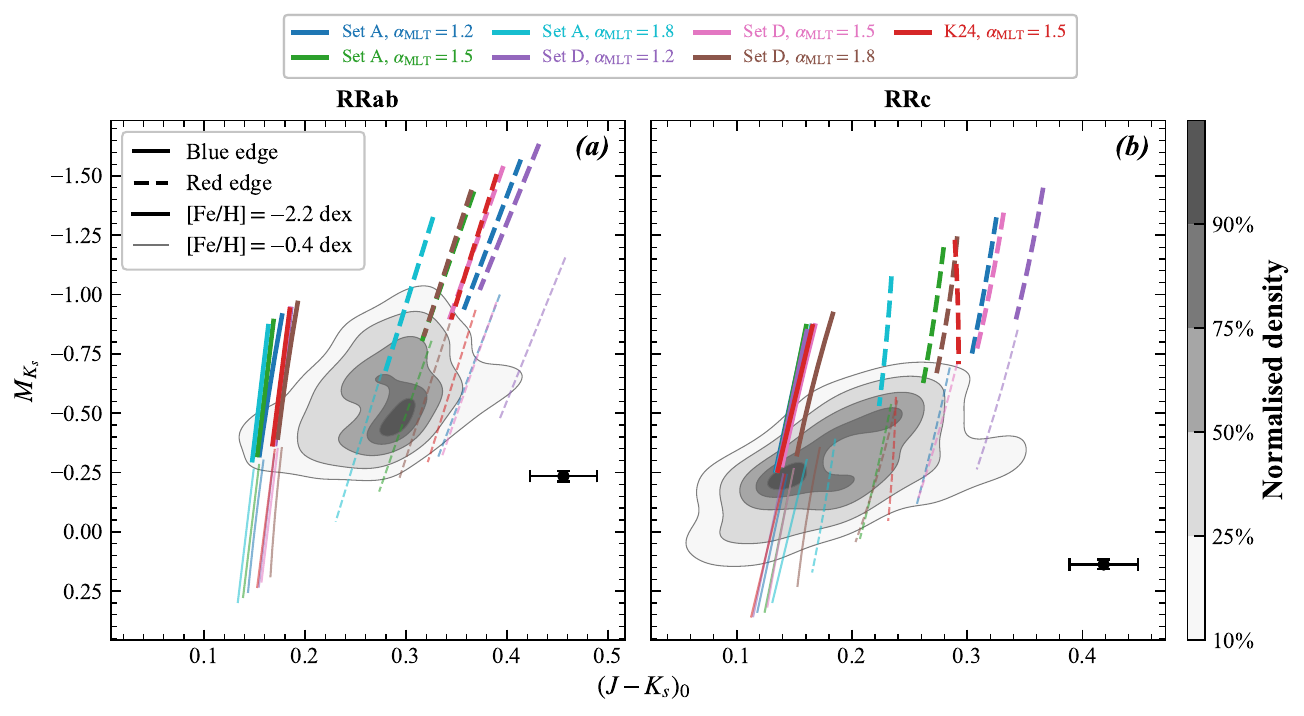}
\caption{Dereddened $(J-K_s)_0$ vs $M_{K_s}$ CMD for the observed
    RRab (panel~a) and RRc (panel~b) samples from \citet{bhar23a}.
    Grey contours show the normalised KDE of the observed distribution
    at the 10, 25, 50, 75, and 90 per cent levels.
    colored solid and dashed lines mark the predicted blue and red edges
    of the instability strip for each convection set at
    $[\mathrm{Fe/H}] = -2.2$\,dex (thick) and $-0.4$\,dex (thin).
    The black error bar in the bottom-right corner represents the
    median photometric uncertainty of the whole sample.}
\label{fig:cmd}
\end{figure}

Tables~\ref{tab:IS_relations_FU} and \ref{tab:IS_relations_FO} provide the coefficients of the IS edges in the $L$--$T_{\rm eff}$ plane for RRab and RRc stars, respectively. In Fig.~\ref{fig:mlt_comp}, we show the comparison of IS edges as a function of $\alpha_{\rm MLT}$ with different convection prescriptions for two representative metallicities. Fig.~\ref{fig:IS_width} shows the IS width as a function of all convection and metallicity prescriptions.  In Fig.~\ref{fig:M15_comp}, we present the comparison of the theoretical IS boundaries for four representative metallicities with different convection prescriptions as well as that of \citetalias{marc15}. Fig.~\ref{fig:stellingwerf_comparison} shows the comparison of IS edges for RRab (left) and RRc (right) at two representative metallicities obtained with three model sets: \citetalias{marc15} (non-linear Stellingwerf model, previous BaSTI ZAHB), \mesarsp\ (linear, updated BaSTI
ZAHB), and new non-linear Stellingwerf models (updated BaSTI ZAHB as in \mesarsp\ case). Fig.~\ref{fig:IS_in_cmd} represents the IS edges in some example filters (\textit{Gaia}, LSST and Keil diagram). Fig.~\ref{fig:cmd} shows the predicted IS edges for all convection sets overlaid on the observed $(J-K_s)_0$ vs $M_{K_s}$ CMD. 

\begin{table*}
\centering
\small
\caption{Coefficients of the IS boundary relations for RRab
models.}
\label{tab:IS_relations_FU}
\setlength{\tabcolsep}{2pt}
\renewcommand{\arraystretch}{1.0}
\begin{tabular}{llclccccccc}
\hline\hline
Set & $\alpha_{\rm mlt}$ & Edge & Form &
$\alpha$ & $\beta$ & $\beta_{2}$ & $\gamma$ &
$R^{2}$ & $\Delta$AIC & $p_{F}$ \\
\hline
\multirow{12}{*}{A}
 & \multirow{4}{*}{1.2}
   & \multirow{2}{*}{Blue}
     & Lin
     & $3.90539 \pm 0.00373$ & $-0.04819 \pm 0.00283$ & $-$
     & $-0.00902 \pm 0.00057$ & $0.872$
     & \multirow{2}{*}{$-21.8$} & \multirow{2}{*}{$<0.001$} \\
 & & & \textbf{Quad}
     & $\mathbf{3.77038 \pm 0.02555}$ & $\mathbf{+0.11616 \pm 0.03097}$
     & $\mathbf{-0.04977 \pm 0.00936}$ & $\mathbf{-0.00910 \pm 0.00045}$
     & $\mathbf{0.922}$ & & \\[3pt]
 & & \multirow{2}{*}{Red}
     & \textbf{Lin}
     & $\mathbf{3.89813 \pm 0.00537}$ & $\mathbf{-0.10660 \pm 0.00408}$
     & $-$ & $\mathbf{-0.01668 \pm 0.00082}$ & $\mathbf{0.938}$
     & \multirow{2}{*}{$+1.9$} & \multirow{2}{*}{$0.794$} \\
 & & & Quad
     & $3.91042 \pm 0.04713$ & $-0.12155 \pm 0.05711$
     & $+0.00453 \pm 0.01724$ & $-0.01667 \pm 0.00083$ & $0.938$ & & \\[6pt]
 & \multirow{4}{*}{1.5}
   & \multirow{2}{*}{Blue}
     & Lin
     & $3.90151 \pm 0.00335$ & $-0.04307 \pm 0.00255$ & $-$
     & $-0.00802 \pm 0.00051$ & $0.870$
     & \multirow{2}{*}{$-3.9$} & \multirow{2}{*}{$0.021$} \\
 & & & \textbf{Quad}
     & $\mathbf{3.83567 \pm 0.02769}$ & $\mathbf{+0.03708 \pm 0.03357}$
     & $\mathbf{-0.02427 \pm 0.01014}$ & $\mathbf{-0.00805 \pm 0.00049}$
     & $\mathbf{0.885}$ & & \\[3pt]
 & & \multirow{2}{*}{Red}
     & \textbf{Lin}
     & $\mathbf{3.91985 \pm 0.00342}$ & $\mathbf{-0.10080 \pm 0.00260}$
     & $-$ & $\mathbf{-0.01211 \pm 0.00052}$ & $\mathbf{0.973}$
     & \multirow{2}{*}{$+2.0$} & \multirow{2}{*}{$0.912$} \\
 & & & Quad
     & $3.92318 \pm 0.03005$ & $-0.10486 \pm 0.03641$
     & $+0.00123 \pm 0.01099$ & $-0.01211 \pm 0.00053$ & $0.973$ & & \\[6pt]
 & \multirow{4}{*}{1.8}
   & \multirow{2}{*}{Blue}
     & Lin
     & $3.90363 \pm 0.00369$ & $-0.04236 \pm 0.00280$ & $-$
     & $-0.00767 \pm 0.00056$ & $0.841$
     & \multirow{2}{*}{$-2.3$} & \multirow{2}{*}{$0.047$} \\
 & & & \textbf{Quad}
     & $\mathbf{3.84084 \pm 0.03096}$ & $\mathbf{+0.03407 \pm 0.03753}$
     & $\mathbf{-0.02314 \pm 0.01134}$ & $\mathbf{-0.00770 \pm 0.00055}$
     & $\mathbf{0.854}$ & & \\[3pt]
 & & \multirow{2}{*}{Red}
     & \textbf{Lin}
     & $\mathbf{3.93522 \pm 0.00315}$ & $\mathbf{-0.09575 \pm 0.00239}$
     & $-$ & $\mathbf{-0.00915 \pm 0.00048}$ & $\mathbf{0.978}$
     & \multirow{2}{*}{$-1.8$} & \multirow{2}{*}{$0.064$} \\
 & & & Quad
     & $3.98554 \pm 0.02661$ & $-0.15697 \pm 0.03225$
     & $+0.01853 \pm 0.00974$ & $-0.00912 \pm 0.00047$ & $0.979$ & & \\
\hline
\multirow{12}{*}{D}
 & \multirow{4}{*}{1.2}
   & \multirow{2}{*}{Blue}
     & \textbf{Lin}
     & $\mathbf{3.89114 \pm 0.00676}$ & $\mathbf{-0.04109 \pm 0.00513}$
     & $-$ & $\mathbf{-0.00780 \pm 0.00104}$ & $\mathbf{0.604}$
     & \multirow{2}{*}{$+0.8$} & \multirow{2}{*}{$0.302$} \\
 & & & Quad
     & $3.83036 \pm 0.05856$ & $+0.03290 \pm 0.07099$
     & $-0.02240 \pm 0.02144$ & $-0.00784 \pm 0.00104$ & $0.613$ & & \\[3pt]
 & & \multirow{2}{*}{Red}
     & \textbf{Lin}
     & $\mathbf{3.85789 \pm 0.01030}$ & $\mathbf{-0.10467 \pm 0.00782}$
     & $-$ & $\mathbf{-0.02505 \pm 0.00158}$ & $\mathbf{0.850}$
     & \multirow{2}{*}{$+1.5$} & \multirow{2}{*}{$0.517$} \\
 & & & Quad
     & $3.79949 \pm 0.09008$ & $-0.03363 \pm 0.10914$
     & $-0.02150 \pm 0.03294$ & $-0.02509 \pm 0.00159$ & $0.852$ & & \\[6pt]
 & \multirow{4}{*}{1.5}
   & \multirow{2}{*}{Blue}
     & Lin
     & $3.88986 \pm 0.00487$ & $-0.04292 \pm 0.00370$ & $-$
     & $-0.00927 \pm 0.00075$ & $0.785$
     & \multirow{2}{*}{$-2.5$} & \multirow{2}{*}{$0.044$} \\
 & & & \textbf{Quad}
     & $\mathbf{3.80576 \pm 0.04077}$ & $\mathbf{+0.05945 \pm 0.04942}$
     & $\mathbf{-0.03100 \pm 0.01492}$ & $\mathbf{-0.00932 \pm 0.00072}$
     & $\mathbf{0.804}$ & & \\[3pt]
 & & \multirow{2}{*}{Red}
     & \textbf{Lin}
     & $\mathbf{3.87593 \pm 0.00733}$ & $\mathbf{-0.09523 \pm 0.00557}$
     & $-$ & $\mathbf{-0.01862 \pm 0.00112}$ & $\mathbf{0.877}$
     & \multirow{2}{*}{$+1.7$} & \multirow{2}{*}{$0.621$} \\
 & & & Quad
     & $3.84416 \pm 0.06424$ & $-0.05658 \pm 0.07783$
     & $-0.01170 \pm 0.02349$ & $-0.01864 \pm 0.00113$ & $0.878$ & & \\[6pt]
 & \multirow{4}{*}{1.8}
   & \multirow{2}{*}{Blue}
     & Lin
     & $3.88107 \pm 0.00593$ & $-0.03954 \pm 0.00450$ & $-$
     & $-0.00945 \pm 0.00091$ & $0.709$
     & \multirow{2}{*}{$-13.4$} & \multirow{2}{*}{$<0.001$} \\
 & & & \textbf{Quad}
     & $\mathbf{3.70117 \pm 0.04436}$ & $\mathbf{+0.17941 \pm 0.05377}$
     & $\mathbf{-0.06630 \pm 0.01624}$ & $\mathbf{-0.00956 \pm 0.00078}$
     & $\mathbf{0.789}$ & & \\[3pt]
 & & \multirow{2}{*}{Red}
     & \textbf{Lin}
     & $\mathbf{3.88940 \pm 0.00507}$ & $\mathbf{-0.08784 \pm 0.00385}$
     & $-$ & $\mathbf{-0.01403 \pm 0.00078}$ & $\mathbf{0.921}$
     & \multirow{2}{*}{$+1.3$} & \multirow{2}{*}{$0.413$} \\
 & & & Quad
     & $3.92568 \pm 0.04419$ & $-0.13198 \pm 0.05354$
     & $+0.01336 \pm 0.01616$ & $-0.01401 \pm 0.00078$ & $0.922$ & & \\
\hline
\multirow{4}{*}{\citetalias{kova24}}
 & \multirow{4}{*}{1.5}
   & \multirow{2}{*}{Blue}
     & \textbf{Lin}
     & $\mathbf{3.89621 \pm 0.00417}$ & $\mathbf{-0.04523 \pm 0.00317}$
     & $-$ & $\mathbf{-0.00877 \pm 0.00064}$ & $\mathbf{0.831}$
     & \multirow{2}{*}{$-0.6$} & \multirow{2}{*}{$0.127$} \\
 & & & Quad
     & $3.84106 \pm 0.03567$ & $+0.02191 \pm 0.04325$
     & $-0.02033 \pm 0.01306$ & $-0.00880 \pm 0.00063$ & $0.840$ & & \\[3pt]
 & & \multirow{2}{*}{Red}
     & \textbf{Lin}
     & $\mathbf{3.87441 \pm 0.00375}$ & $\mathbf{-0.08389 \pm 0.00284}$
     & $-$ & $\mathbf{-0.01353 \pm 0.00057}$ & $\mathbf{0.951}$
     & \multirow{2}{*}{$+1.5$} & \multirow{2}{*}{$0.512$} \\
 & & & Quad
     & $3.85290 \pm 0.03274$ & $-0.05772 \pm 0.03967$
     & $-0.00792 \pm 0.01198$ & $-0.01355 \pm 0.00058$ & $0.951$ & & \\
\hline
\end{tabular}
\tablefoot{For each boundary both the linear (Eq.~\ref{eq:IS_linear})
and quadratic (Eq.~\ref{eq:IS_quad}) fits are reported.
$\Delta\mathrm{AIC} = \mathrm{AIC_{quad}} - \mathrm{AIC_{lin}}$ and
$p_{F}$ are from the $F$-test on $\beta_{2}$ and are shared between
the two rows of each boundary. The statistically preferred form
($\Delta\mathrm{AIC}<-2$ and $p_{F}<0.05$) is highlighted in
bold. A dash in $\beta_{2}$ denotes $\beta_{2}\equiv0$.
Uncertainties are $1\sigma$.}
\end{table*}

\begin{table*}
\centering
\small
\caption{Coefficients of the IS boundary relations for RRc
models.}
\label{tab:IS_relations_FO}
\setlength{\tabcolsep}{2pt}
\renewcommand{\arraystretch}{1.1}
\begin{tabular}{llclccccccc}
\hline\hline
Set & $\alpha_{\rm mlt}$ & Edge & Form &
$\alpha$ & $\beta$ & $\beta_{2}$ & $\gamma$ &
$R^{2}$ & $\Delta$AIC & $p_{F}$ \\
\hline
\multirow{12}{*}{A}
 & \multirow{4}{*}{1.2}
   & \multirow{2}{*}{Blue}
     & Lin
     & $3.94492 \pm 0.00328$ & $-0.06907 \pm 0.00249$ & $-$
     & $-0.01082 \pm 0.00050$ & $0.945$
     & \multirow{2}{*}{$-10.3$} & \multirow{2}{*}{$<0.001$} \\
 & & & \textbf{Quad}
     & $\mathbf{3.85467 \pm 0.02532}$ & $\mathbf{+0.04079 \pm 0.03070}$
     & $\mathbf{-0.03327 \pm 0.00927}$ & $\mathbf{-0.01087 \pm 0.00045}$
     & $\mathbf{0.957}$ & & \\[3pt]
 & & \multirow{2}{*}{Red}
     & Lin
     & $3.87186 \pm 0.00315$ & $-0.05412 \pm 0.00239$ & $-$
     & $-0.00442 \pm 0.00048$ & $0.938$
     & \multirow{2}{*}{$-7.4$} & \multirow{2}{*}{$0.003$} \\
 & & & \textbf{Quad}
     & $\mathbf{3.94896 \pm 0.02510}$ & $\mathbf{-0.14795 \pm 0.03042}$
     & $\mathbf{+0.02841 \pm 0.00918}$ & $\mathbf{-0.00437 \pm 0.00044}$
     & $\mathbf{0.949}$ & & \\[6pt]
 & \multirow{4}{*}{1.5}
   & \multirow{2}{*}{Blue}
     & Lin
     & $3.93026 \pm 0.00435$ & $-0.06291 \pm 0.00330$ & $-$
     & $-0.01169 \pm 0.00067$ & $0.895$
     & \multirow{2}{*}{$-7.8$} & \multirow{2}{*}{$0.003$} \\
 & & & \textbf{Quad}
     & $\mathbf{3.82223 \pm 0.03451}$ & $\mathbf{+0.06859 \pm 0.04184}$
     & $\mathbf{-0.03982 \pm 0.01264}$ & $\mathbf{-0.01175 \pm 0.00061}$
     & $\mathbf{0.914}$ & & \\[3pt]
 & & \multirow{2}{*}{Red}
     & Lin
     & $3.89752 \pm 0.00458$ & $-0.05467 \pm 0.00348$ & $-$
     & $-0.00201 \pm 0.00070$ & $0.911$
     & \multirow{2}{*}{$-15.0$} & \multirow{2}{*}{$<0.001$} \\
 & & & \textbf{Quad}
     & $\mathbf{4.04207 \pm 0.03370}$ & $\mathbf{-0.23058 \pm 0.04085}$
     & $\mathbf{+0.05326 \pm 0.01234}$ & $\mathbf{-0.00192 \pm 0.00060}$
     & $\mathbf{0.938}$ & & \\[6pt]
 & \multirow{4}{*}{1.8}
   & \multirow{2}{*}{Blue}
     & \textbf{Lin}
     & $\mathbf{3.94194 \pm 0.00572}$ & $\mathbf{-0.07651 \pm 0.00425}$
     & $-$ & $\mathbf{-0.01487 \pm 0.00088}$ & $\mathbf{0.903}$
     & \multirow{2}{*}{$-2.2$} & \multirow{2}{*}{$0.052$} \\
 & & & Quad
     & $3.82751 \pm 0.05737$ & $+0.06046 \pm 0.06847$
     & $-0.04074 \pm 0.02033$ & $-0.01483 \pm 0.00085$ & $0.913$ & & \\[3pt]
 & & \multirow{2}{*}{Red}
     & Lin
     & $3.90113 \pm 0.00484$ & $-0.04048 \pm 0.00360$ & $-$
     & $+0.00165 \pm 0.00075$ & $0.903$
     & \multirow{2}{*}{$-8.2$} & \multirow{2}{*}{$0.003$} \\
 & & & \textbf{Quad}
     & $\mathbf{4.04634 \pm 0.04510}$ & $\mathbf{-0.21427 \pm 0.05382}$
     & $\mathbf{+0.05169 \pm 0.01598}$ & $\mathbf{+0.00160 \pm 0.00067}$
     & $\mathbf{0.925}$ & & \\
\hline
\multirow{12}{*}{D}
 & \multirow{4}{*}{1.2}
   & \multirow{2}{*}{Blue}
     & \textbf{Lin}
     & $\mathbf{3.94920 \pm 0.00271}$ & $\mathbf{-0.06882 \pm 0.00206}$
     & $-$ & $\mathbf{-0.00932 \pm 0.00042}$ & $\mathbf{0.962}$
     & \multirow{2}{*}{$+0.1$} & \multirow{2}{*}{$0.194$} \\
 & & & Quad
     & $3.91860 \pm 0.02337$ & $-0.03158 \pm 0.02833$
     & $-0.01128 \pm 0.00856$ & $-0.00934 \pm 0.00041$ & $0.964$ & & \\[3pt]
 & & \multirow{2}{*}{Red}
     & Lin
     & $3.84941 \pm 0.00402$ & $-0.05730 \pm 0.00305$ & $-$
     & $-0.00815 \pm 0.00062$ & $0.888$
     & \multirow{2}{*}{$-4.1$} & \multirow{2}{*}{$0.019$} \\
 & & & \textbf{Quad}
     & $\mathbf{3.92986 \pm 0.03315}$ & $\mathbf{-0.15518 \pm 0.04017}$
     & $\mathbf{+0.02963 \pm 0.01213}$ & $\mathbf{-0.00810 \pm 0.00058}$
     & $\mathbf{0.901}$ & & \\[6pt]
 & \multirow{4}{*}{1.5}
   & \multirow{2}{*}{Blue}
     & Lin
     & $3.93354 \pm 0.00402$ & $-0.06409 \pm 0.00305$ & $-$
     & $-0.01071 \pm 0.00061$ & $0.908$
     & \multirow{2}{*}{$-13.9$} & \multirow{2}{*}{$<0.001$} \\
 & & & \textbf{Quad}
     & $\mathbf{3.81001 \pm 0.02991}$ & $\mathbf{+0.08627 \pm 0.03625}$
     & $\mathbf{-0.04553 \pm 0.01095}$ & $\mathbf{-0.01078 \pm 0.00053}$
     & $\mathbf{0.934}$ & & \\[3pt]
 & & \multirow{2}{*}{Red}
     & Lin
     & $3.88603 \pm 0.00377$ & $-0.06500 \pm 0.00286$ & $-$
     & $-0.00536 \pm 0.00058$ & $0.939$
     & \multirow{2}{*}{$-18.5$} & \multirow{2}{*}{$<0.001$} \\
 & & & \textbf{Quad}
     & $\mathbf{4.01467 \pm 0.02676}$ & $\mathbf{-0.22152 \pm 0.03243}$
     & $\mathbf{+0.04738 \pm 0.00979}$ & $\mathbf{-0.00528 \pm 0.00047}$
     & $\mathbf{0.960}$ & & \\[6pt]
 & \multirow{4}{*}{1.8}
   & \multirow{2}{*}{Blue}
     & Lin
     & $3.91555 \pm 0.00686$ & $-0.06271 \pm 0.00505$ & $-$
     & $-0.01330 \pm 0.00102$ & $0.828$
     & \multirow{2}{*}{$-11.7$} & \multirow{2}{*}{$<0.001$} \\
 & & & \textbf{Quad}
     & $\mathbf{3.68222 \pm 0.06128}$ & $\mathbf{+0.21650 \pm 0.07312}$
     & $\mathbf{-0.08314 \pm 0.02173}$ & $\mathbf{-0.01334 \pm 0.00088}$
     & $\mathbf{0.876}$ & & \\[3pt]
 & & \multirow{2}{*}{Red}
     & Lin
     & $3.90893 \pm 0.00632$ & $-0.06149 \pm 0.00465$ & $-$
     & $-0.00107 \pm 0.00094$ & $0.892$
     & \multirow{2}{*}{$-17.0$} & \multirow{2}{*}{$<0.001$} \\
 & & & \textbf{Quad}
     & $\mathbf{4.15464 \pm 0.05294}$ & $\mathbf{-0.35543 \pm 0.06315}$
     & $\mathbf{+0.08750 \pm 0.01878}$ & $\mathbf{-0.00103 \pm 0.00076}$
     & $\mathbf{0.931}$ & & \\
\hline
\multirow{4}{*}{\citetalias{kova24}}
 & \multirow{4}{*}{1.5}
   & \multirow{2}{*}{Blue}
     & Lin
     & $3.96268 \pm 0.00308$ & $-0.07839 \pm 0.00234$ & $-$
     & $-0.01011 \pm 0.00047$ & $0.963$
     & \multirow{2}{*}{$-10.0$} & \multirow{2}{*}{$<0.001$} \\
 & & & \textbf{Quad}
     & $\mathbf{3.87874 \pm 0.02389}$ & $\mathbf{+0.02381 \pm 0.02897}$
     & $\mathbf{-0.03095 \pm 0.00875}$ & $\mathbf{-0.01016 \pm 0.00042}$
     & $\mathbf{0.971}$ & & \\[3pt]
 & & \multirow{2}{*}{Red}
     & Lin
     & $3.83693 \pm 0.00583$ & $-0.01072 \pm 0.00443$ & $-$
     & $+0.00542 \pm 0.00089$ & $0.784$
     & \multirow{2}{*}{$-4.1$} & \multirow{2}{*}{$0.018$} \\
 & & & \textbf{Quad}
     & $\mathbf{3.95384 \pm 0.04799}$ & $\mathbf{-0.15300 \pm 0.05817}$
     & $\mathbf{+0.04308 \pm 0.01757}$ & $\mathbf{+0.00549 \pm 0.00085}$
     & $\mathbf{0.810}$ & & \\
\hline
\end{tabular}
\tablefoot{Same as Table~\ref{tab:IS_relations_FU}. The RRc grids
at $\alpha_{\rm mlt}=1.8$ use 41 data points for Set~A and 42 for
Set~D, owing to the absence of unstable pulsation models at
$Z=0.024270$ and $Z=0.032580$. The relations should be applied
within the luminosity ranges appropriate for each metallicity, as
listed in Table~\ref{tab:logL_ranges}.}
\end{table*}

\section{PLZ/PWZ relations}
\begin{figure}
\centering
\includegraphics[width=0.95\textwidth,keepaspectratio]
{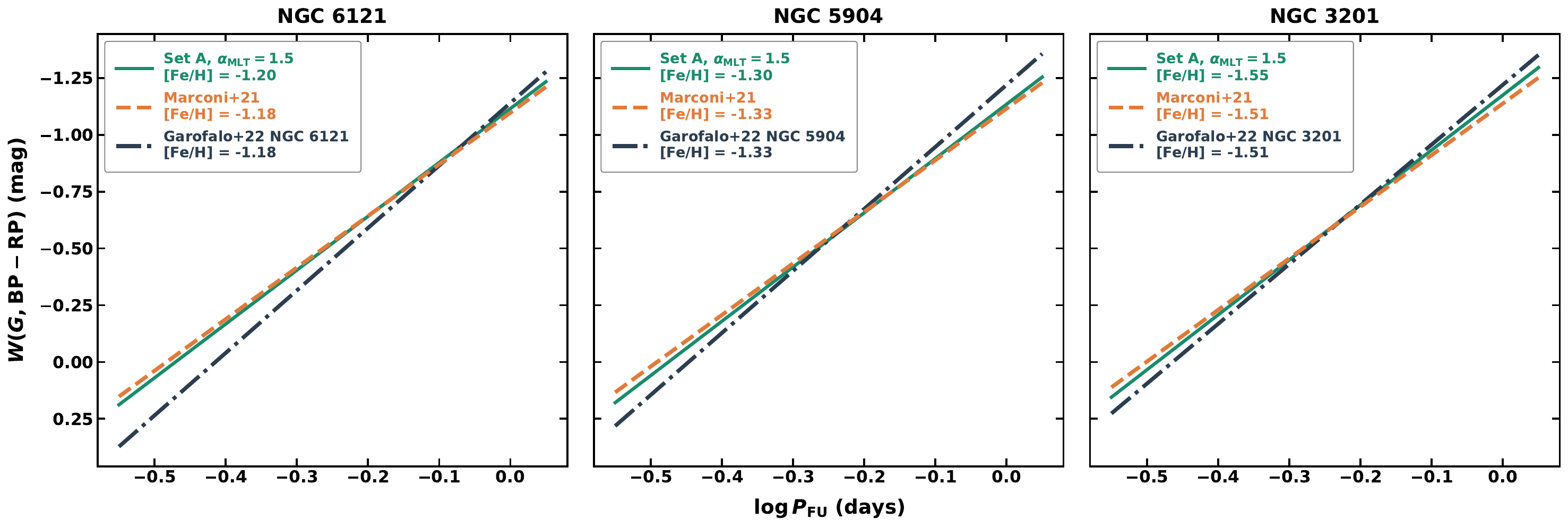}
\caption{Comparison of the \textit{Gaia} PW relations from the present work (the relation with Set A, $\alpha_{\mathrm{MLT}} = 1.5$; best matched convection prescription with the empirical relation in this case) with the theoretical relations of \citet{marc21} and the empirical relations of \citet{garo22}, for three individual globular clusters with metallicities close to our grid values: NGC~6121 (left), NGC~5904 (middle), and NGC~3201 (right). The corresponding metallicities adopted for each relation are given in the legends.}
\label{fig:gaia_pwz}
\end{figure}

In this section, we present the PLZ/PWZ relations in the NIR bands: Table~\ref{tab:best_fit} gives the best-matched relation compared with that of \citet{bhar23a}, and Table~\ref{tab:plz_combined} gives the relations for the whole theoretical grid. Tables~\ref{tab:pwz_gaia_coeffs} and \ref{tab:gaia_pw_relations} present the PWZ relations in the \textit{Gaia} bands for Milky Way field and globular cluster RRLs, respectively. Fig.~\ref{fig:gaia_pwz} presents the \textit{Gaia} PWZ relation for three globular clusters and a comparison with those of \citet{marc21,garo22}.

\begin{table*}
\centering
\caption{Best-matching predicted PLZ and PWZ relations for RRab, RRc, and
combined (RRab$+$RRc) samples in the $JHK_s$ bands, restricted to
$[\mathrm{Fe/H}] < -0.4$\,dex.}
\label{tab:best_fit}
\setlength{\tabcolsep}{4pt}
\begin{tabular}{lllrrrrrrrr}
\toprule
Type & Band & Best-matching set
  & $a$ & $\sigma_a$
  & $b$ & $\sigma_b$
  & $c$ & $\sigma_c$
  & $\sigma$ & $N$ \\
\midrule
\multirow{6}{*}{RRab}
 & $J$                & Set\,D, $\alpha_\mathrm{MLT}=1.2$
   & $-0.461$ & $0.009$ & $-2.010$ & $0.027$ & $+0.219$ & $0.005$ & $0.085$ &   943 \\
 & $H$                & Set\,A, $\alpha_\mathrm{MLT}=1.2$
   & $-0.736$ & $0.005$ & $-2.353$ & $0.014$ & $+0.215$ & $0.003$ & $0.048$ &  1027 \\
 & $K_s$              &  \citetalias{kova24}
   & $-0.804$ & $0.005$ & $-2.431$ & $0.014$ & $+0.211$ & $0.003$ & $0.042$ &   908 \\
 & $W_{J,H}$          & Set\,D, $\alpha_\mathrm{MLT}=1.2$
   & $-1.244$ & $0.005$ & $-3.013$ & $0.017$ & $+0.206$ & $0.003$ & $0.052$ &   943 \\
 & $W_{J,K_s}$        & Set\,D, $\alpha_\mathrm{MLT}=1.2$
   & $-1.014$ & $0.003$ & $-2.645$ & $0.010$ & $+0.206$ & $0.002$ & $0.030$ &   943 \\
 & $W_{H,K_s}$        &  \citetalias{kova24}
   & $-0.899$ & $0.004$ & $-2.466$ & $0.013$ & $+0.206$ & $0.003$ & $0.040$ &   908 \\
\midrule
\multirow{6}{*}{RRc}
 & $J$                & Set\,D, $\alpha_\mathrm{MLT}=1.2$
   & $-0.570$ & $0.015$ & $-2.110$ & $0.039$ & $+0.188$ & $0.006$ & $0.082$ &   622 \\
 & $H$                & Set\,D, $\alpha_\mathrm{MLT}=1.2$
   & $-0.819$ & $0.009$ & $-2.423$ & $0.023$ & $+0.187$ & $0.004$ & $0.048$ &   622 \\
 & $K_s$              & Set\,D, $\alpha_\mathrm{MLT}=1.2$
   & $-0.869$ & $0.008$ & $-2.445$ & $0.022$ & $+0.185$ & $0.004$ & $0.046$ &   622 \\
 & $W_{J,H}$          & Set\,A, $\alpha_\mathrm{MLT}=1.5$
   & $-1.226$ & $0.006$ & $-2.905$ & $0.015$ & $+0.193$ & $0.002$ & $0.028$ &   508 \\
 & $W_{J,K_s}$        & Set\,D, $\alpha_\mathrm{MLT}=1.2$
   & $-1.076$ & $0.005$ & $-2.676$ & $0.013$ & $+0.182$ & $0.002$ & $0.027$ &   622 \\
 & $W_{H,K_s}$        & Set\,D, $\alpha_\mathrm{MLT}=1.2$
   & $-0.966$ & $0.008$ & $-2.486$ & $0.021$ & $+0.179$ & $0.003$ & $0.044$ &   622 \\
\midrule
\multirow{6}{*}{RRab$+$RRc}
 & $J$                & Set\,D, $\alpha_\mathrm{MLT}=1.2$
   & $-0.475$ & $0.007$ & $-1.934$ & $0.020$ & $+0.208$ & $0.004$ & $0.087$ &  1565 \\
 & $H$                & Set\,D, $\alpha_\mathrm{MLT}=1.2$
   & $-0.749$ & $0.004$ & $-2.303$ & $0.011$ & $+0.205$ & $0.002$ & $0.050$ &  1565 \\
 & $K_s$              &  \citetalias{kova24}
   & $-0.818$ & $0.004$ & $-2.372$ & $0.011$ & $+0.201$ & $0.002$ & $0.046$ &  1476 \\
 & $W_{J,H}$          & Set\,D, $\alpha_\mathrm{MLT}=1.2$
   & $-1.251$ & $0.004$ & $-2.978$ & $0.011$ & $+0.198$ & $0.002$ & $0.047$ &  1565 \\
 & $W_{J,K_s}$        & Set\,D, $\alpha_\mathrm{MLT}=1.2$
   & $-1.025$ & $0.003$ & $-2.597$ & $0.007$ & $+0.198$ & $0.002$ & $0.032$ &  1565 \\
 & $W_{H,K_s}$        & \citetalias{kova24}
   & $-0.913$ & $0.004$ & $-2.410$ & $0.010$ & $+0.195$ & $0.002$ & $0.044$ &  1476 \\
\bottomrule
\end{tabular}
\tablefoot{For each pulsation type and band the best model is selected by minimising
the normalised Euclidean distance in $(a,b,c)$ space to the reference
relation: Bhardwaj et al.\ (2023) for RRab and RRab$+$RRc, and
Marconi et al.\ (2015, FO) for RRc.
The relation has the form
$M_\lambda = a + b\log P + c\,[\mathrm{Fe/H}]$,
where $P$ is the fundamental-mode period (RRab and combined) or the
first-overtone period (RRc).
RRc periods in the combined sample are fundamentalised via
$\log P_{\rm FU} = \log P_{\rm FO} + 0.127$ (Bhardwaj et al.\ 2023).
Wesenheit magnitudes:
$W_{J,H}=H-1.83(J-H)$,
$W_{J,K_s}=K_s-0.69(J-K_s)$,
$W_{H,K_s}=K_s-1.92(H-K_s)$.}
\end{table*}

\begin{table*}
\centering
\caption{Predicted RRab, RRc, and combined (RRab$+$RRc) PLZ and PWZ relations
($M_\lambda = a + b\log P + c\,[\mathrm{Fe/H}]$) in the $JHK_s$ bands for
different convection sets and mixing-length parameters $\alpha_\mathrm{MLT}$,
shown side by side for the three pulsation-type samples.}
\label{tab:plz_combined}
\resizebox{\textwidth}{!}{%
\setlength{\tabcolsep}{4pt}
\begin{tabular}{ll ccc rr ccc rr ccc rr}
\toprule
& & \multicolumn{5}{c}{RRab} & \multicolumn{5}{c}{RRc} & \multicolumn{5}{c}{RRab$+$RRc} \\
\cmidrule(lr){3-7} \cmidrule(lr){8-12} \cmidrule(lr){13-17}
Set & Band
  & $a$ & $b$ & $c$ & $\sigma$ & $N$
  & $a$ & $b$ & $c$ & $\sigma$ & $N$
  & $a$ & $b$ & $c$ & $\sigma$ & $N$ \\
\midrule
\multirow{6}{*}{Set\,A, $\alpha_\mathrm{MLT}=1.2$}
 & $J$        & $-0.478\pm0.008$ & $-2.021\pm0.024$ & $+0.216\pm0.005$ & $0.082$ & 1027
              & $-0.626\pm0.013$ & $-2.258\pm0.036$ & $+0.184\pm0.006$ & $0.073$ &  616
              & $-0.498\pm0.007$ & $-1.986\pm0.018$ & $+0.205\pm0.004$ & $0.082$ & 1643 \\
 & $H$        & $-0.736\pm0.005$ & $-2.353\pm0.014$ & $+0.215\pm0.003$ & $0.048$ & 1027
              & $-0.847\pm0.008$ & $-2.504\pm0.022$ & $+0.186\pm0.003$ & $0.045$ &  616
              & $-0.753\pm0.004$ & $-2.318\pm0.011$ & $+0.205\pm0.002$ & $0.050$ & 1643 \\
 & $K_s$      & $-0.786\pm0.005$ & $-2.373\pm0.014$ & $+0.212\pm0.003$ & $0.046$ & 1027
              & $-0.897\pm0.008$ & $-2.523\pm0.021$ & $+0.183\pm0.003$ & $0.044$ &  616
              & $-0.803\pm0.004$ & $-2.339\pm0.011$ & $+0.202\pm0.002$ & $0.049$ & 1643 \\
 & $W_{J,H}$  & $-1.210\pm0.005$ & $-2.961\pm0.014$ & $+0.213\pm0.003$ & $0.046$ & 1027
              & $-1.253\pm0.006$ & $-2.954\pm0.015$ & $+0.189\pm0.002$ & $0.031$ &  616
              & $-1.219\pm0.004$ & $-2.927\pm0.009$ & $+0.204\pm0.002$ & $0.042$ & 1643 \\
 & $W_{J,K_s}$& $-0.999\pm0.003$ & $-2.616\pm0.009$ & $+0.209\pm0.002$ & $0.031$ & 1027
              & $-1.083\pm0.005$ & $-2.705\pm0.014$ & $+0.182\pm0.002$ & $0.028$ &  616
              & $-1.013\pm0.003$ & $-2.583\pm0.007$ & $+0.200\pm0.002$ & $0.033$ & 1643 \\
 & $W_{H,K_s}$& $-0.882\pm0.004$ & $-2.412\pm0.013$ & $+0.207\pm0.003$ & $0.044$ & 1027
              & $-0.991\pm0.008$ & $-2.558\pm0.020$ & $+0.177\pm0.003$ & $0.042$ &  616
              & $-0.898\pm0.004$ & $-2.379\pm0.010$ & $+0.197\pm0.002$ & $0.046$ & 1643 \\
\midrule
\multirow{6}{*}{Set\,A, $\alpha_\mathrm{MLT}=1.5$}
 & $J$        & $-0.509\pm0.008$ & $-2.103\pm0.022$ & $+0.216\pm0.005$ & $0.072$ &  928
              & $-0.691\pm0.012$ & $-2.406\pm0.031$ & $+0.177\pm0.005$ & $0.059$ &  508
              & $-0.534\pm0.007$ & $-2.079\pm0.017$ & $+0.204\pm0.004$ & $0.073$ & 1436 \\
 & $H$        & $-0.744\pm0.005$ & $-2.378\pm0.014$ & $+0.216\pm0.003$ & $0.045$ &  928
              & $-0.880\pm0.008$ & $-2.583\pm0.021$ & $+0.183\pm0.003$ & $0.039$ &  508
              & $-0.764\pm0.004$ & $-2.355\pm0.011$ & $+0.206\pm0.002$ & $0.048$ & 1436 \\
 & $K_s$      & $-0.793\pm0.005$ & $-2.395\pm0.013$ & $+0.214\pm0.003$ & $0.044$ &  928
              & $-0.928\pm0.008$ & $-2.598\pm0.020$ & $+0.180\pm0.003$ & $0.038$ &  508
              & $-0.813\pm0.004$ & $-2.373\pm0.011$ & $+0.203\pm0.002$ & $0.047$ & 1436 \\
 & $W_{J,H}$  & $-1.175\pm0.004$ & $-2.882\pm0.013$ & $+0.218\pm0.003$ & $0.041$ &  928
              & $-1.226\pm0.006$ & $-2.905\pm0.015$ & $+0.193\pm0.002$ & $0.028$ &  508
              & $-1.185\pm0.003$ & $-2.860\pm0.009$ & $+0.209\pm0.002$ & $0.038$ & 1436 \\
 & $W_{J,K_s}$& $-0.989\pm0.004$ & $-2.597\pm0.010$ & $+0.212\pm0.002$ & $0.032$ &  928
              & $-1.091\pm0.006$ & $-2.731\pm0.015$ & $+0.182\pm0.002$ & $0.028$ &  508
              & $-1.005\pm0.003$ & $-2.575\pm0.008$ & $+0.203\pm0.002$ & $0.034$ & 1436 \\
 & $W_{H,K_s}$& $-0.887\pm0.005$ & $-2.428\pm0.013$ & $+0.209\pm0.003$ & $0.042$ &  928
              & $-1.020\pm0.007$ & $-2.629\pm0.019$ & $+0.175\pm0.003$ & $0.036$ &  508
              & $-0.906\pm0.004$ & $-2.406\pm0.011$ & $+0.198\pm0.002$ & $0.045$ & 1436 \\
\midrule
\multirow{6}{*}{Set\,A, $\alpha_\mathrm{MLT}=1.8$}
 & $J$        & $-0.538\pm0.008$ & $-2.136\pm0.020$ & $+0.210\pm0.004$ & $0.064$ &  831
              & $-0.756\pm0.012$ & $-2.544\pm0.028$ & $+0.172\pm0.005$ & $0.041$ &  313
              & $-0.554\pm0.007$ & $-2.126\pm0.017$ & $+0.205\pm0.004$ & $0.065$ & 1144 \\
 & $H$        & $-0.751\pm0.005$ & $-2.378\pm0.014$ & $+0.215\pm0.003$ & $0.043$ &  831
              & $-0.921\pm0.009$ & $-2.672\pm0.021$ & $+0.179\pm0.003$ & $0.030$ &  313
              & $-0.765\pm0.005$ & $-2.368\pm0.012$ & $+0.209\pm0.003$ & $0.045$ & 1144 \\
 & $K_s$      & $-0.799\pm0.005$ & $-2.394\pm0.013$ & $+0.212\pm0.003$ & $0.043$ &  831
              & $-0.968\pm0.008$ & $-2.686\pm0.020$ & $+0.177\pm0.003$ & $0.030$ &  313
              & $-0.813\pm0.005$ & $-2.384\pm0.012$ & $+0.206\pm0.003$ & $0.044$ & 1144 \\
 & $W_{J,H}$  & $-1.142\pm0.005$ & $-2.822\pm0.012$ & $+0.223\pm0.003$ & $0.038$ &  831
              & $-1.223\pm0.007$ & $-2.904\pm0.016$ & $+0.192\pm0.003$ & $0.023$ &  313
              & $-1.152\pm0.004$ & $-2.810\pm0.009$ & $+0.216\pm0.002$ & $0.036$ & 1144 \\
 & $W_{J,K_s}$& $-0.979\pm0.004$ & $-2.572\pm0.011$ & $+0.214\pm0.002$ & $0.033$ &  831
              & $-1.114\pm0.007$ & $-2.784\pm0.016$ & $+0.180\pm0.003$ & $0.024$ &  313
              & $-0.992\pm0.004$ & $-2.562\pm0.009$ & $+0.207\pm0.002$ & $0.035$ & 1144 \\
 & $W_{H,K_s}$& $-0.891\pm0.005$ & $-2.424\pm0.013$ & $+0.207\pm0.003$ & $0.041$ &  831
              & $-1.058\pm0.008$ & $-2.715\pm0.019$ & $+0.172\pm0.003$ & $0.029$ &  313
              & $-0.905\pm0.005$ & $-2.416\pm0.011$ & $+0.202\pm0.003$ & $0.043$ & 1144 \\
\midrule
\multirow{6}{*}{Set\,D, $\alpha_\mathrm{MLT}=1.2$}
 & $J$        & $-0.461\pm0.009$ & $-2.010\pm0.027$ & $+0.219\pm0.005$ & $0.085$ &  943
              & $-0.570\pm0.015$ & $-2.110\pm0.039$ & $+0.188\pm0.006$ & $0.082$ &  622
              & $-0.475\pm0.007$ & $-1.934\pm0.020$ & $+0.208\pm0.004$ & $0.087$ & 1565 \\
 & $H$        & $-0.737\pm0.005$ & $-2.364\pm0.015$ & $+0.215\pm0.003$ & $0.047$ &  943
              & $-0.819\pm0.009$ & $-2.423\pm0.023$ & $+0.187\pm0.004$ & $0.048$ &  622
              & $-0.749\pm0.004$ & $-2.303\pm0.011$ & $+0.205\pm0.002$ & $0.050$ & 1565 \\
 & $K_s$      & $-0.788\pm0.005$ & $-2.386\pm0.015$ & $+0.212\pm0.003$ & $0.046$ &  943
              & $-0.869\pm0.008$ & $-2.445\pm0.022$ & $+0.185\pm0.004$ & $0.046$ &  622
              & $-0.800\pm0.004$ & $-2.326\pm0.011$ & $+0.202\pm0.002$ & $0.049$ & 1565 \\
 & $W_{J,H}$  & $-1.244\pm0.005$ & $-3.013\pm0.017$ & $+0.206\pm0.003$ & $0.052$ &  943
              & $-1.275\pm0.007$ & $-2.996\pm0.017$ & $+0.186\pm0.003$ & $0.036$ &  622
              & $-1.251\pm0.004$ & $-2.978\pm0.011$ & $+0.198\pm0.002$ & $0.047$ & 1565 \\
 & $W_{J,K_s}$& $-1.014\pm0.003$ & $-2.645\pm0.010$ & $+0.206\pm0.002$ & $0.030$ &  943
              & $-1.076\pm0.005$ & $-2.676\pm0.013$ & $+0.182\pm0.002$ & $0.027$ &  622
              & $-1.025\pm0.003$ & $-2.597\pm0.007$ & $+0.198\pm0.002$ & $0.032$ & 1565 \\
 & $W_{H,K_s}$& $-0.886\pm0.004$ & $-2.427\pm0.014$ & $+0.206\pm0.003$ & $0.043$ &  943
              & $-0.966\pm0.008$ & $-2.486\pm0.021$ & $+0.179\pm0.003$ & $0.044$ &  622
              & $-0.898\pm0.004$ & $-2.370\pm0.010$ & $+0.196\pm0.002$ & $0.046$ & 1565 \\
\midrule
\multirow{6}{*}{Set\,D, $\alpha_\mathrm{MLT}=1.5$}
 & $J$        & $-0.494\pm0.008$ & $-2.103\pm0.025$ & $+0.215\pm0.005$ & $0.076$ &  857
              & $-0.628\pm0.014$ & $-2.278\pm0.039$ & $+0.183\pm0.006$ & $0.072$ &  537
              & $-0.511\pm0.007$ & $-2.040\pm0.019$ & $+0.204\pm0.004$ & $0.078$ & 1394 \\
 & $H$        & $-0.751\pm0.005$ & $-2.406\pm0.015$ & $+0.214\pm0.003$ & $0.044$ &  857
              & $-0.850\pm0.009$ & $-2.519\pm0.023$ & $+0.186\pm0.004$ & $0.044$ &  537
              & $-0.765\pm0.004$ & $-2.356\pm0.012$ & $+0.204\pm0.002$ & $0.048$ & 1394 \\
 & $K_s$      & $-0.800\pm0.005$ & $-2.424\pm0.014$ & $+0.211\pm0.003$ & $0.043$ &  857
              & $-0.900\pm0.008$ & $-2.537\pm0.023$ & $+0.183\pm0.003$ & $0.042$ &  537
              & $-0.815\pm0.004$ & $-2.375\pm0.011$ & $+0.201\pm0.002$ & $0.046$ & 1394 \\
 & $W_{J,H}$  & $-1.219\pm0.005$ & $-2.960\pm0.016$ & $+0.212\pm0.003$ & $0.047$ &  857
              & $-1.258\pm0.006$ & $-2.962\pm0.017$ & $+0.191\pm0.003$ & $0.032$ &  537
              & $-1.228\pm0.004$ & $-2.933\pm0.010$ & $+0.204\pm0.002$ & $0.042$ & 1394 \\
 & $W_{J,K_s}$& $-1.011\pm0.003$ & $-2.645\pm0.010$ & $+0.209\pm0.002$ & $0.031$ &  857
              & $-1.087\pm0.005$ & $-2.717\pm0.015$ & $+0.183\pm0.002$ & $0.027$ &  537
              & $-1.024\pm0.003$ & $-2.606\pm0.008$ & $+0.200\pm0.002$ & $0.032$ & 1394 \\
 & $W_{H,K_s}$& $-0.896\pm0.005$ & $-2.459\pm0.014$ & $+0.206\pm0.003$ & $0.041$ &  857
              & $-0.994\pm0.008$ & $-2.572\pm0.022$ & $+0.178\pm0.003$ & $0.040$ &  537
              & $-0.910\pm0.004$ & $-2.412\pm0.011$ & $+0.196\pm0.002$ & $0.044$ & 1394 \\
\midrule
\multirow{6}{*}{Set\,D, $\alpha_\mathrm{MLT}=1.8$}
 & $J$        & $-0.521\pm0.008$ & $-2.190\pm0.024$ & $+0.213\pm0.005$ & $0.065$ &  729
              & $-0.684\pm0.014$ & $-2.437\pm0.037$ & $+0.174\pm0.006$ & $0.053$ &  349
              & $-0.537\pm0.007$ & $-2.138\pm0.019$ & $+0.204\pm0.004$ & $0.066$ & 1078 \\
 & $H$        & $-0.760\pm0.005$ & $-2.442\pm0.015$ & $+0.214\pm0.003$ & $0.040$ &  729
              & $-0.885\pm0.009$ & $-2.625\pm0.024$ & $+0.182\pm0.004$ & $0.035$ &  349
              & $-0.773\pm0.004$ & $-2.403\pm0.012$ & $+0.206\pm0.003$ & $0.043$ & 1078 \\
 & $K_s$      & $-0.809\pm0.005$ & $-2.458\pm0.015$ & $+0.212\pm0.003$ & $0.040$ &  729
              & $-0.934\pm0.009$ & $-2.641\pm0.024$ & $+0.179\pm0.004$ & $0.034$ &  349
              & $-0.823\pm0.004$ & $-2.419\pm0.012$ & $+0.203\pm0.002$ & $0.042$ & 1078 \\
 & $W_{J,H}$  & $-1.198\pm0.006$ & $-2.905\pm0.017$ & $+0.217\pm0.003$ & $0.045$ &  729
              & $-1.254\pm0.006$ & $-2.968\pm0.016$ & $+0.196\pm0.002$ & $0.023$ &  349
              & $-1.206\pm0.004$ & $-2.888\pm0.011$ & $+0.211\pm0.002$ & $0.040$ & 1078 \\
 & $W_{J,K_s}$& $-1.008\pm0.004$ & $-2.642\pm0.011$ & $+0.211\pm0.002$ & $0.031$ &  729
              & $-1.107\pm0.006$ & $-2.781\pm0.017$ & $+0.182\pm0.002$ & $0.024$ &  349
              & $-1.020\pm0.003$ & $-2.614\pm0.009$ & $+0.203\pm0.002$ & $0.032$ & 1078 \\
 & $W_{H,K_s}$& $-0.903\pm0.005$ & $-2.486\pm0.014$ & $+0.207\pm0.003$ & $0.038$ &  729
              & $-1.028\pm0.008$ & $-2.671\pm0.023$ & $+0.173\pm0.003$ & $0.033$ &  349
              & $-0.917\pm0.004$ & $-2.451\pm0.012$ & $+0.198\pm0.002$ & $0.041$ & 1078 \\
\midrule
\multirow{6}{*}{\citetalias{kova24}}
 & $J$        & $-0.502\pm0.008$ & $-2.117\pm0.024$ & $+0.213\pm0.005$ & $0.074$ &  908
              & $-0.662\pm0.013$ & $-2.317\pm0.033$ & $+0.176\pm0.005$ & $0.066$ &  568
              & $-0.519\pm0.007$ & $-2.034\pm0.018$ & $+0.202\pm0.004$ & $0.077$ & 1476 \\
 & $H$        & $-0.754\pm0.005$ & $-2.414\pm0.014$ & $+0.213\pm0.003$ & $0.043$ &  908
              & $-0.870\pm0.008$ & $-2.539\pm0.021$ & $+0.181\pm0.003$ & $0.041$ &  568
              & $-0.768\pm0.004$ & $-2.352\pm0.011$ & $+0.203\pm0.002$ & $0.047$ & 1476 \\
 & $K_s$      & $-0.804\pm0.005$ & $-2.431\pm0.014$ & $+0.211\pm0.003$ & $0.042$ &  908
              & $-0.919\pm0.008$ & $-2.557\pm0.020$ & $+0.179\pm0.003$ & $0.040$ &  568
              & $-0.818\pm0.004$ & $-2.372\pm0.011$ & $+0.201\pm0.002$ & $0.046$ & 1476 \\
 & $W_{J,H}$  & $-1.217\pm0.005$ & $-2.957\pm0.015$ & $+0.214\pm0.003$ & $0.047$ &  908
              & $-1.250\pm0.006$ & $-2.944\pm0.014$ & $+0.191\pm0.002$ & $0.029$ &  568
              & $-1.225\pm0.004$ & $-2.933\pm0.010$ & $+0.206\pm0.002$ & $0.042$ & 1476 \\
 & $W_{J,K_s}$& $-1.012\pm0.003$ & $-2.649\pm0.010$ & $+0.209\pm0.002$ & $0.031$ &  908
              & $-1.096\pm0.005$ & $-2.723\pm0.013$ & $+0.181\pm0.002$ & $0.027$ &  568
              & $-1.025\pm0.003$ & $-2.605\pm0.008$ & $+0.200\pm0.002$ & $0.033$ & 1476 \\
 & $W_{H,K_s}$& $-0.899\pm0.004$ & $-2.466\pm0.013$ & $+0.206\pm0.003$ & $0.040$ &  908
              & $-1.012\pm0.007$ & $-2.592\pm0.019$ & $+0.174\pm0.003$ & $0.038$ &  568
              & $-0.913\pm0.004$ & $-2.410\pm0.010$ & $+0.195\pm0.002$ & $0.044$ & 1476 \\
\bottomrule
\end{tabular}%
}
\tablefoot{Wesenheit magnitudes are defined as $W_{J,H}=H-1.83(J-H)$,
$W_{J,K_s}=K_s-0.69(J-K_s)$, and $W_{H,K_s}=K_s-1.92(H-K_s)$.
\rev{RRab stars are selected by $\dot{\gamma}_\mathrm{FU}>0$ and $0.29\leq P\leq1.00$\,d; RRc by $\dot{\gamma}_\mathrm{FU}>0$ and $0.20\leq P\leq0.51$\,d.}
Columns are grouped by pulsation-type sample: RRab, RRc, and the combined
RRab$+$RRc sample (same convection set/$\alpha_\mathrm{MLT}$ combination in
each row).}
\end{table*}
\begin{table*}
\centering
\caption{PWZ relation coefficients in $W(G,\,\mathrm{BP-RP}) = a + b\log P_{\rm FU} + c\,\mathrm{[Fe/H]}$ for RRab, RRc, and the combined (GLOBAL) samples, across all convection prescriptions.}
\label{tab:pwz_gaia_coeffs}
\begin{tabular}{lccccc}
\hline\hline
Model & $a$ & $b$ & $c$ & $\sigma$ (mag) & $N$ \\
\hline
\multicolumn{6}{c}{\textit{RRab }} \\
Set A, $\alpha_\mathrm{MLT}=1.2$ & $-0.949 \pm 0.005$ & $-2.421 \pm 0.015$ & $+0.149 \pm 0.002$ & 0.047 & 1144 \\
Set A, $\alpha_\mathrm{MLT}=1.5$ & $-0.961 \pm 0.005$ & $-2.456 \pm 0.015$ & $+0.151 \pm 0.002$ & 0.044 & 1044 \\
Set A, $\alpha_\mathrm{MLT}=1.8$ & $-0.967 \pm 0.005$ & $-2.469 \pm 0.014$ & $+0.153 \pm 0.002$ & 0.041 & 926 \\
Set D, $\alpha_\mathrm{MLT}=1.2$ & $-0.949 \pm 0.005$ & $-2.417 \pm 0.016$ & $+0.148 \pm 0.002$ & 0.047 & 1044 \\
Set D, $\alpha_\mathrm{MLT}=1.5$ & $-0.965 \pm 0.005$ & $-2.470 \pm 0.016$ & $+0.147 \pm 0.002$ & 0.045 & 953 \\
Set D, $\alpha_\mathrm{MLT}=1.8$ & $-0.979 \pm 0.005$ & $-2.513 \pm 0.016$ & $+0.147 \pm 0.002$ & 0.041 & 802 \\
K24 & $-0.970 \pm 0.005$ & $-2.479 \pm 0.015$ & $+0.147 \pm 0.002$ & 0.044 & 1006 \\
\hline
\multicolumn{6}{c}{\textit{RRc}} \\
Set A, $\alpha_\mathrm{MLT}=1.2$ & $-1.381 \pm 0.016$ & $-2.535 \pm 0.032$ & $+0.114 \pm 0.003$ & 0.038 & 486 \\
Set A, $\alpha_\mathrm{MLT}=1.5$ & $-1.418 \pm 0.017$ & $-2.600 \pm 0.034$ & $+0.109 \pm 0.003$ & 0.036 & 387 \\
Set A, $\alpha_\mathrm{MLT}=1.8$ & $-1.464 \pm 0.019$ & $-2.732 \pm 0.036$ & $+0.126 \pm 0.003$ & 0.027 & 207 \\
Set D, $\alpha_\mathrm{MLT}=1.2$ & $-1.358 \pm 0.016$ & $-2.492 \pm 0.033$ & $+0.117 \pm 0.003$ & 0.039 & 487 \\
Set D, $\alpha_\mathrm{MLT}=1.5$ & $-1.394 \pm 0.017$ & $-2.559 \pm 0.034$ & $+0.113 \pm 0.003$ & 0.038 & 421 \\
Set D, $\alpha_\mathrm{MLT}=1.8$ & $-1.406 \pm 0.022$ & $-2.611 \pm 0.043$ & $+0.118 \pm 0.004$ & 0.031 & 221 \\
K24 & $-1.398 \pm 0.016$ & $-2.551 \pm 0.031$ & $+0.109 \pm 0.003$ & 0.037 & 463 \\
\hline
\multicolumn{6}{c}{\textit{Global (RRab + fund. RRc)}} \\
Set A, $\alpha_\mathrm{MLT}=1.2$ & $-0.948 \pm 0.004$ & $-2.339 \pm 0.012$ & $+0.140 \pm 0.002$ & 0.049 & 1630 \\
Set A, $\alpha_\mathrm{MLT}=1.5$ & $-0.961 \pm 0.004$ & $-2.390 \pm 0.012$ & $+0.143 \pm 0.002$ & 0.047 & 1431 \\
Set A, $\alpha_\mathrm{MLT}=1.8$ & $-0.964 \pm 0.004$ & $-2.443 \pm 0.012$ & $+0.153 \pm 0.002$ & 0.041 & 1133 \\
Set D, $\alpha_\mathrm{MLT}=1.2$ & $-0.942 \pm 0.004$ & $-2.314 \pm 0.012$ & $+0.140 \pm 0.002$ & 0.049 & 1531 \\
Set D, $\alpha_\mathrm{MLT}=1.5$ & $-0.959 \pm 0.004$ & $-2.366 \pm 0.013$ & $+0.139 \pm 0.002$ & 0.047 & 1374 \\
Set D, $\alpha_\mathrm{MLT}=1.8$ & $-0.970 \pm 0.004$ & $-2.450 \pm 0.013$ & $+0.146 \pm 0.002$ & 0.041 & 1023 \\
K24 & $-0.962 \pm 0.004$ & $-2.365 \pm 0.012$ & $+0.138 \pm 0.002$ & 0.047 & 1469 \\
\hline
\end{tabular}
\end{table*}
\begin{table*}
\centering
\caption{Fitted $PW(G,\mathrm{BP-RP})$ relations, $W = a + b\log P_{\rm FU}$, for each convective set / mixing-length prescription, evaluated over the observed period range of RRab and RRc stars in each \citet{garo22} globular cluster.}
\label{tab:gaia_pw_relations}
\resizebox{\textwidth}{!}{
\begin{tabular}{l ccc c ccc c ccc}
\hline\hline
 & \multicolumn{3}{c}{NGC 6121 ([Fe/H]=$-1.18$)} & & \multicolumn{3}{c}{NGC 5904 ([Fe/H]=$-1.33$)} & & \multicolumn{3}{c}{NGC 3201 ([Fe/H]=$-1.51$)} \\
\cline{2-4} \cline{6-8} \cline{10-12}
Set & $a$ & $b$ & $\sigma$ ($N$) & & $a$ & $b$ & $\sigma$ ($N$) & & $a$ & $b$ & $\sigma$ ($N$) \\
\hline
Set A, $\alpha_{\rm MLT}=1.2$ & $-1.101 \pm 0.013$ & $-2.321 \pm 0.053$ & 0.049 (81) & & $-1.118 \pm 0.014$ & $-2.338 \pm 0.054$ & 0.048 (91) & & $-1.158 \pm 0.027$ & $-2.359 \pm 0.090$ & 0.045 (47) \\
Set A, $\alpha_{\rm MLT}=1.5$ & $-1.115 \pm 0.014$ & $-2.370 \pm 0.055$ & 0.046 (73) & & $-1.135 \pm 0.013$ & $-2.388 \pm 0.053$ & 0.045 (82) & & $-1.175 \pm 0.026$ & $-2.418 \pm 0.090$ & 0.045 (49) \\
Set A, $\alpha_{\rm MLT}=1.8$ & $-1.126 \pm 0.013$ & $-2.393 \pm 0.054$ & 0.042 (62) & & $-1.150 \pm 0.013$ & $-2.430 \pm 0.053$ & 0.041 (71) & & $-1.213 \pm 0.027$ & $-2.536 \pm 0.094$ & 0.043 (43) \\
Set D, $\alpha_{\rm MLT}=1.2$ & $-1.093 \pm 0.013$ & $-2.287 \pm 0.055$ & 0.049 (76) & & $-1.111 \pm 0.013$ & $-2.303 \pm 0.054$ & 0.048 (85) & & $-1.160 \pm 0.031$ & $-2.360 \pm 0.106$ & 0.044 (37) \\
Set D, $\alpha_{\rm MLT}=1.5$ & $-1.113 \pm 0.014$ & $-2.352 \pm 0.058$ & 0.047 (65) & & $-1.130 \pm 0.014$ & $-2.382 \pm 0.057$ & 0.045 (72) & & $-1.165 \pm 0.027$ & $-2.385 \pm 0.093$ & 0.042 (38) \\
Set D, $\alpha_{\rm MLT}=1.8$ & $-1.124 \pm 0.013$ & $-2.407 \pm 0.055$ & 0.040 (57) & & $-1.142 \pm 0.013$ & $-2.423 \pm 0.055$ & 0.040 (64) & & $-1.158 \pm 0.030$ & $-2.378 \pm 0.105$ & 0.040 (33) \\
K24 & $-1.115 \pm 0.013$ & $-2.359 \pm 0.050$ & 0.045 (75) & & $-1.133 \pm 0.013$ & $-2.382 \pm 0.051$ & 0.044 (84) & & $-1.152 \pm 0.026$ & $-2.341 \pm 0.086$ & 0.042 (44) \\
\hline
\end{tabular}
}
\end{table*}

\end{appendix}

\end{document}